\documentclass[aps,prb,notitlepage,reprint,onecolumn,showpacs,amsmath,amssymb,superscriptaddress,longbibliography,10pt]{revtex4-2}

\usepackage{graphicx}
\usepackage{dcolumn}
\usepackage{bm}

\usepackage{braket}
\usepackage{natbib}
\usepackage{amsmath}
\usepackage{mathtools}

\usepackage{hyperref}
\hypersetup{pdfstartview={FitH},pdfpagemode={UseNone},
	colorlinks,linkcolor=blue, citecolor=blue, urlcolor=blue,
	bookmarksopen=true, pdfnewwindow=true}
\usepackage[all]{hypcap}

\begin{document}

\title{Observation of arbitrary topological windings of a non-Hermitian band}

\author{Kai Wang}
\thanks{These authors contributed equally}
 \affiliation{Ginzton Laboratory and Department of Electrical Engineering, Stanford University, Stanford, CA 94305, USA.}

\author{Avik Dutt}
\thanks{These authors contributed equally}
\affiliation{Ginzton Laboratory and Department of Electrical Engineering, Stanford University, Stanford, CA 94305, USA.}

\author{Ki Youl Yang}
\affiliation{Ginzton Laboratory and Department of Electrical Engineering, Stanford University, Stanford, CA 94305, USA.}

\author{Casey~C.~Wojcik}
\affiliation{Ginzton Laboratory and Department of Electrical Engineering, Stanford University, Stanford, CA 94305, USA.}

\author{Jelena Vu\v{c}kovi\'c}
\affiliation{Ginzton Laboratory and Department of Electrical Engineering, Stanford University, Stanford, CA 94305, USA.}

\author{Shanhui Fan}
\email{shanhui@stanford.edu}
\affiliation{Ginzton Laboratory and Department of Electrical Engineering, Stanford University, Stanford, CA 94305, USA.}


\begin{abstract}

The non-trivial topological features in the energy band of non-Hermitian systems provide promising pathways to achieve robust physical behaviors in classical or quantum open systems. A key topological feature, unique to non-Hermitian systems, is the non-trivial winding of the energy band in the complex energy plane. Here we provide direct experimental demonstrations of such non-trivial winding, by implementing non-Hermitian lattice Hamiltonians along a frequency synthetic dimension formed in a ring resonator undergoing simultaneous phase and amplitude modulations, and by directly characterizing the complex band structures. Moreover, we show that the topological winding can be straightforwardly controlled by changing the modulation waveform. Our results open a pathway for the experimental synthesis and characterization of topologically non-trivial phases in non-conservative systems.   
\end{abstract}

\maketitle

The discoveries of a wide variety of topological materials, with topological insulators~\cite{kane_Z2_2005, bernevig_quantum_2006, hsieh_topological_2008, konig_quantum_2007} and Weyl semimetals~\cite{lu_experimental_2015, xu_discovery_2015} being two prominent examples, highlight the importance of topological band theory, which seeks to develop fundamental understandings of topological properties of energy band structures. Early efforts on topological band theory have focused on Hermitian Hamiltonians~\cite{qi_topological_2011}. In recent years, motivated in part by the significant developments in topological photonics~\cite{wang_observation_2009, rechtsman_photonic_2013, hafezi_imaging_2013, fang_realizing_2012, ozawa_topological_2019-1}, where gain and loss are common features~\cite{Weimann2017,Bandres2018, zhao_non-hermitian_2019, bahari_nonreciprocal_2017}, there have been significant emerging interests in developing topological band theory for non-Hermitian Hamiltonians~\cite{shen_topological_2018,Gong2018,Kawabata2019,Wojcik2020}. 

The energy bands of non-Hermitian Hamiltonians exhibit unique non-trivial topological features that are absent in Hermitian systems. In particular, since the energy bands of non-Hermitian Hamiltonians are in general complex, generically even a single energy band in one dimension can form a non-trivial loop in the complex plane, as characterized by integer non-zero winding numbers~\cite{Gong2018, Wojcik2020}. This is in contrast with Hermitian systems, where non-trivial topology requires at least two bands, and moreover requires symmetry protection in one dimension~\cite{su_solitons_1979}. Such non-trivial winding, which is unique to non-Hermitian systems, provides the topological underpinnings of remarkable phenomena such as the non-Hermitian skin effect~\cite{Weidemann2020,Xiao2020,Lee2016,Yao2018,Longhi2019a,Borgnia2020,Okuma2020} and necessitates the generalization of bulk-edge correspondence~\cite{Kunst2018,Yokomizo2019,Imura2019}.  

In spite of the central importance of energy band winding in the topological band theory of non-Hermitian systems, direct experimental observation and control of such winding have been lacking. Recent experiments have demonstrated the non-Hermitian skin effect~\cite{Weidemann2020,Xiao2020, Ghatak2020}, as well as the collapse of eigenspectrum as induced by the presence of an edge~\cite{Helbig2020}. These experiments, however, only provide \emph{indirect} evidence of energy-band winding in non-Hermitian systems.  Moreover, strictly speaking, energy bands are defined only for infinite systems. For non-Hermitian systems, the eigenstates of a finite lattice can be qualitatively different from those in an infinite lattice. Thus, it is not \emph{a priori} obvious how one can measure the band structure in experimentally feasible non-Hermitian lattices which are typically finite.

In this work, we experimentally implement a class of non-Hermitian Hamiltonians and measure its momentum-resolved complex band energy. Our implementation utilizes the concept of a synthetic dimension~\cite{boada_quantum_2012, celi_synthetic_2014,Yuan2018, Ozawa2019, Lustig2019}. In particular, we use a synthetic frequency dimension~\cite{Dutt2020,Yuan2016, Ozawa2016, Yuan2018a, Dutt2020a, buddhiraju_arbitrary_2020, Wang2020} as formed by multiple frequency modes in a ring resonator under amplitude and phase modulations. Our measurement provides direct visualization of non-trivial topological band winding. Moreover, we demonstrate that the winding can be controlled by varying the parameters of the modulation, and highly complex winding as characterized by multiple different non-zero winding numbers can be achieved.

Specifically, in this paper we implement in the frequency synthetic dimension the 1D lattice Hamiltonian
\begin{equation}~\label{eq:hamil}
    \mathbf{H}=\sum_{m,n} \left (\kappa_{+m}\mathbf{a_{n+m}^\dagger}\mathbf{a_{n}}
    +  \kappa_{-m}\mathbf{a_{n}^\dagger}\mathbf{a_{n+m}}\right) ,
\end{equation}
where $\mathbf{a_{n}^\dagger}$ ($\mathbf{a_{n}}$) is the creation (annihilation) operator of the $n$-th lattice site with $m=1,2,\dots,M$ running over the coupling orders.
This Hamiltonian becomes non-Hermitian if $\kappa_{+m}\neq \kappa_{-m}^\ast$ for some $m$ in Eq.~\eqref{eq:hamil}. 
A general non-Hermitian form of the coupling constant $\kappa_{\pm m}$ can be given by $\kappa_{\pm m}=C_{m} e^{\pm i\alpha_m} \pm \Delta_{m} e^{\pm i\beta_m}$ with the coupling strengths $C_m, \Delta_m \geq 0$ and phases $\alpha_m,\beta_m\in (-\pi,\pi]$, where $C_{m} e^{\pm i\alpha_m}$ is the Hermitian part and $\pm \Delta_{m} e^{\pm i\beta_m}$ is the anti-Hermitian part.
The Hamiltonian of Eq.~\eqref{eq:hamil} has been extensively explored theoretically~\cite{Hatano1996,Gong2018, wanjura_topological_2020}; however, it is challenging to implement such a Hamiltonian experimentally. The usual implementation of non-Hermitian Hamiltonians through the use of space-dependent gain/loss ~\cite{Wimmer2015,Weimann2017,Bandres2018, zhao_non-hermitian_2019} does not straightforwardly create the non-Hermitian coupling in Eq.~\eqref{eq:hamil}. 
Non-Hermitian couplings were only realized very recently on platforms including pulse trains~\cite{Weidemann2020} and electronic circuits~\cite{Helbig2020}; these demonstrations however implemented only a limited subset of non-Hermitian coupling, i.e. neareast-neighbor coupling (exclusively $m=1$) with restricted in-phase Hermitian and anti-Hermitian parts ($\alpha_1=\beta_1$).  It remains an open question if one can realize general non-Hermitian couplings, which include a control over coupling strengths ($C_m,\Delta_m$), phases ($\alpha_m,\beta_m$) and long coupling ranges ($m>1$)~\cite{Bell2017,  Dutt2019, Hu2020,Gonzalez-Tudela2015}. 
Our results in this paper indicate the significant potential of the platform of synthetic dimension to explore non-Hermitian topological physics and to implement Hamiltonians that are difficult to achieve in other means.

\begin{figure}[t]
\includegraphics[width=0.6\columnwidth]{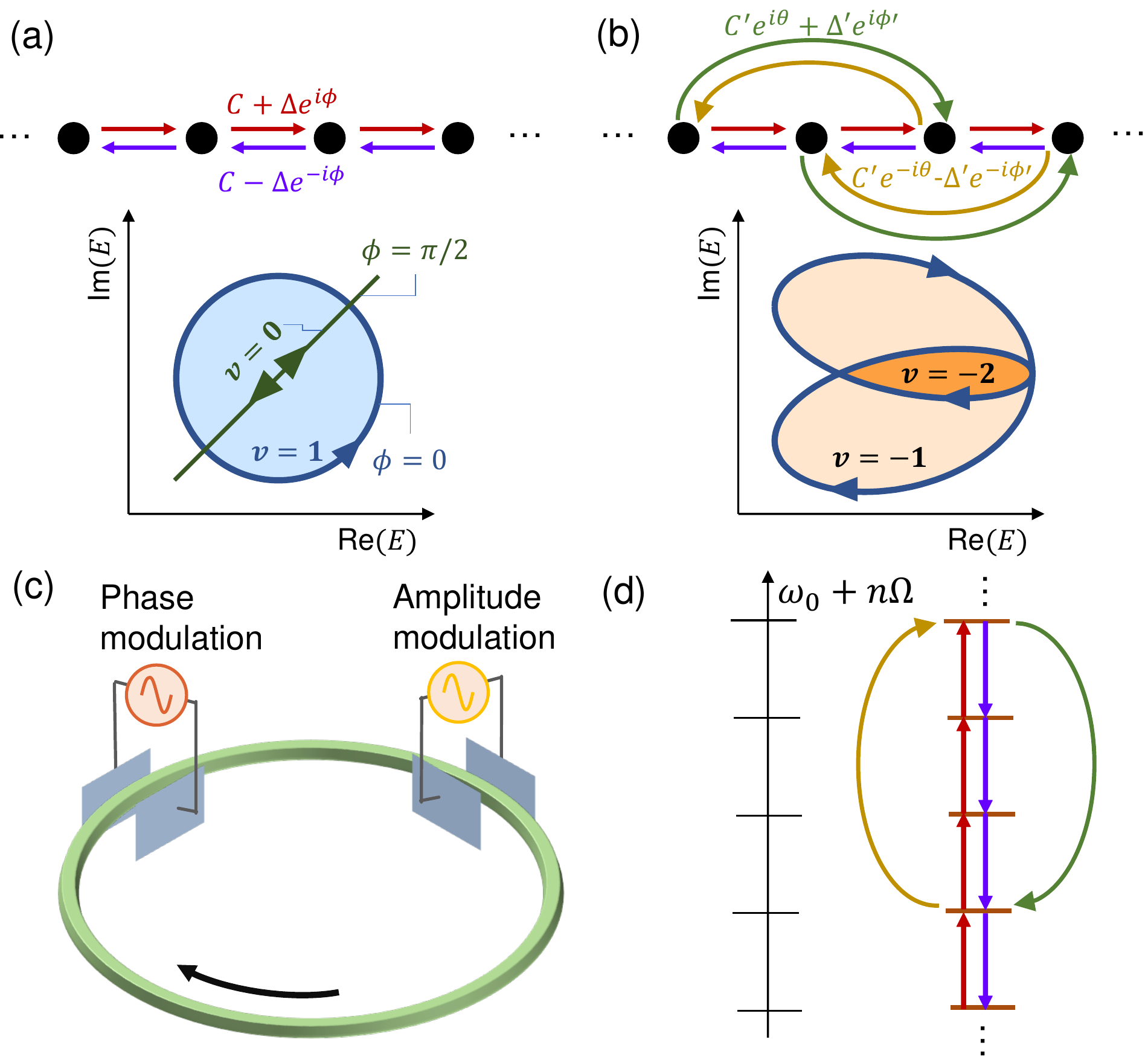}
\caption{{\bf Topological winding numbers in a non-Hermitian 1D lattice and its realization in a synthetic frequency dimension.} (a) Top: non-Hermitian 1D lattice with nearest-neighbor non-Hermitian coupling. 
Bottom: The corresponding band energy winding in the complex plane, as the quasimomentum $k$ goes from $0$ to $2\pi$. 
(b) Same as (a) but with long range coupling. A representative example of $v=-2$ is realized when a reference energy $\epsilon$ is in the inner dark-orange shaded region of the complex plane. (c) Realization of non-Hermitian 1D lattices using a ring resonator undergoing simultaneous amplitude and phase modulation at integer multiples of the free-spectral range $\Omega$. (d)~Left: frequency modes  separated by $\Omega$ (FSR) in the unmodulated resonator. Right: non-Hermitian coupling among the frequency modes created by both phase and amplitude modulation, shown for 1st and 3rd order couplings as an illustrative example.
}
\label{fig:1_abstract}
\end{figure}

The Hamiltonian of Eq.~\eqref{eq:hamil} exhibits a rich set of non-trivial topological behaviors that depend on the strength, phase and range of the coupling. 
We first consider the case with only nearest-neighbor coupling, where the only non-zero coupling constants are $\kappa_{-1} = C - \Delta \exp{(-i\phi)}$ and $\kappa_{+1} = C+\Delta \exp{(i\phi)}$ with $C,\Delta \in \mathbb{R}$, where the 1st term $C$ describes the Hermitian part and the 2nd term $\pm \Delta \exp{(\pm i\phi)}$ is the anti-Hermitian part.
The special case with $\phi=0$ gives rise to the well-known Hatano-Nelson model~\cite{Hatano1996}, where the band winds along an ellipse in the complex plane, as shown in Fig.~\ref{fig:1_abstract}(a,bottom). For a reference energy $\epsilon \in \mathbb{C}$, a winding number of such a 1D band can be defined by~\cite{Okuma2020,Gong2018,Kawabata2019}
\begin{equation}\label{eq:wn}
    v\coloneqq \int_{0}^{2\pi} \frac{{\rm d} k}{2\pi i} \frac{\rm d}{\mathrm{d}k}\ln\,  \left[E(k)-\epsilon\right].
\end{equation}
For the Hatano-Nelson model [Fig.~\ref{fig:1_abstract}(a)], if the reference energy $\epsilon$ is in the interior of the winding loop, e.g. in the shaded area of Fig.~\ref{fig:1_abstract}(a,bottom), one obtains $v=1$.  We note that such a  non-trivial winding is a topological feature that is unique to non-Hermitian Hamiltonians. For a Hermitian Hamiltonian, $E(k)$ is restricted to the real axis, and $v = 0$. 
We also note that the phase difference $\phi$ between the Hermitian and non-Hermitian parts of the coupling can change the shape of the loop. In the special example of $\phi=\pi/2$, $E(k)$ is restricted to a line and hence $v=0$, as shown by the green line in Fig.~\ref{fig:1_abstract}(a,bottom).
More complex winding of the band structure with $|v| \ne 0,1$ can be achieved by introducing long-range coupling into the non-Hermitian Hamiltonian. As an example, in Fig.~\ref{fig:1_abstract}(b), we consider the case with $C=0$, $\Delta/C'= \Delta'/C' = 1$, $\phi=\pi$, $\theta=0$, and $\phi'=\pi$, and $\phi'=3\pi/4$, where $v = -2$ for a suitable choice of the reference energy. In general, 
the sign of $v$ is determined by the handedness of the winding, and $|v|$ depends on how many times the reference point $\epsilon$ is enclosed by the loop in each orientation.

\begin{figure*}[t]
    \centering
    \includegraphics[width=1\textwidth]{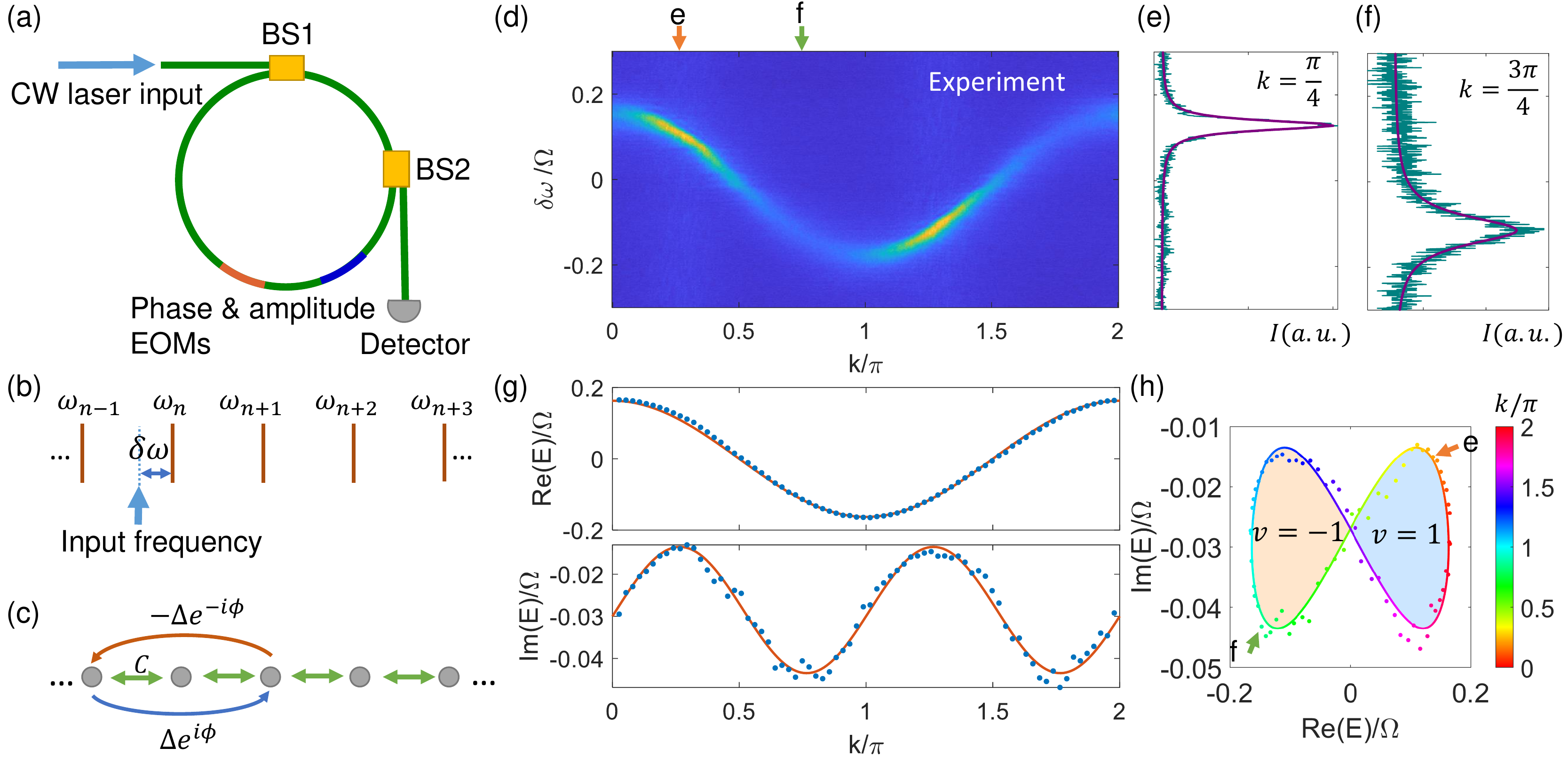}
    \caption{{\bf Measurement of the $k$-resolved complex band energy in synthetic frequency space.} (a)~Input and output schematic of an optical fiber resonator simultaneously modulated by phase and amplitude electro-optic modulators (EOMs). (b) A diagram of several resonant modes of the unmodulated resonator separated by the FSR $\Omega$, where the input laser frequency (arrow) is detuned by $\delta \omega$ from one of the resonance frequencies. (c)~An example of the formed non-Hermitian lattice.  (d)~Color plot of experimentally measured output intensity $I(k,\delta \omega)$. (e,f)~Measurement signal from two example vertical slices (teal curve) and Lorentzian fitting (purple curve) with (e) $k=\pi/4$ and (f) $k=3\pi/4$. (g)~Extracted $k$-dependent $\mathrm{Re}(E)$ and $\mathrm{Im}(E)$ (dots) from (d), compared to the theoretically predicted form (curve). (h)~The winding of the measured energy in the complex plane. Arrows indicate the corresponding points of the $k$-slices shown in (e,f). }
    \label{fig2}
\end{figure*}

To experimentally realize the Hamiltonian of Eq.~\eqref{eq:hamil}, we utilize the synthetic space as formed by the multiple frequency modes of a ring resonator.  
In the absence of modulation, the ring supports a set of resonant modes propagating along a specific direction [e.g. clockwise as shown in Fig.~\ref{fig:1_abstract}(c)]. These modes have frequencies $\omega_n = \omega_0 + n \Omega$, where $n$ is an integer that indexes the mode. $\Omega$ is the free spectral range (FSR), as shown in Fig.~\ref{fig:1_abstract}(d,left).  
Hermitian coupling between nearest-neighbor frequency modes can be formed with the use of a phase modulation having an associated amplitude transmission factor $T_{Ph}=\exp{(-i\delta \Phi)}$ with $\delta \Phi(t)= A_1 \cos \Omega t$. The strength of Hermitian coupling $C \propto A_1$. Similarly, anti-Hermitian coupling can be formed with the use of an amplitude modulation having a transmission factor $T_{Am}=1+B_1 \sin (\Omega t+\phi)$. This gives rise to an anti-Hermitian coupling $\kappa_{\pm 1}=\pm \Delta \exp (\pm i\phi)$ with the strength $\Delta \propto B_1$. Here $\phi$ readily gives rise to the phase difference between the Hermitian and anti-Hermitian parts of the coupling.
With simultaneous phase and amplitude modulations, the equations of motion are given by
\begin{equation}
     \frac{\mathrm{d}}{\mathrm{d} \tau} a_n= -i( C +\Delta e^{i\phi}) a_{n-1}-i ( C-\Delta e^{-i\phi} ) a_{n+1},
\end{equation}
where $a_n$ is the electric-field amplitude of the $n$-th frequency mode and $\tau$ is a slow-time variable counting the number of round trips that light has propagated in the resonator. This realizes the model of Fig.~\ref{fig:1_abstract}(a).
Longer-range coupling can be achieved with modulators having modulation frequencies $m\Omega$ with $m>1$. Specifically, 
simultaneous implementation of the phase modulation $T_{Ph}=\exp{[-i\sum_m A_m \cos (m\Omega t+\alpha_m)]}$ and the amplitude modulation $T_{Am}=1+\sum_m B_m \sin (m\Omega t+\beta_m)$ will give rise to the Hamiltonian in Eq.~\eqref{eq:hamil} 
where 
$\kappa_{\pm m}=C_{m} e^{\pm i\alpha_m} \pm \Delta_{m} e^{\pm i\beta_m}$ (for more details see Sec.~S1A of the supplementary).
Consequently, this method allows for the realization of an arbitrary complex band structure for such one-band 1D non-Hermitian lattices, 
as illustrated in Fig.~\ref{fig:1_abstract}(d) for a specific example. We note that the ability to simultaneously incorporate phase and amplitude modulations is essential for demonstrating topological winding. Hamiltonians with phase or amplitude modulation~\cite{Yuan2018b} alone do not exhibit non-trivial winding in their energy bands.

Now we illustrate how we directly measure the $k$-resolved non-Hermitian band structure in synthetic frequency space. 
As illustrated in Fig.~\ref{fig2}(a), we use a fiber ring resonator undergoing both phase and amplitude modulations by electro-optic modulators (EOMs). We launch continuous-wave (CW) laser light into the ring via a beam splitter (BS1). A second beam splitter BS2 samples a small portion of the intracavity light for time-resolved detection (for more details on the experimental setup see Sec.~S2 of the supplementary). The input laser frequency can be swept across a certain resonance of the unmodulated resonator to access various detunings $\delta \omega$, as illustrated in Fig.~\ref{fig2}(b). Since the lattice space is formed by frequency modes, its reciprocal space is inherently time $t$~\cite{Dutt2019}, with a round-trip time equivalent to one Brillouin zone; hence we can define $k=t \Omega $ as the quasimomentum. The detected signal $I(k,\delta \omega)$ can be related to the Green's function, $G(k,\delta \omega) =i [\delta \omega-E(k)]^{-1}$, as
\begin{equation}\label{eq:lorentzian}
    I(k,\delta \omega)\propto |G(k,\delta \omega)|^2=\frac{1}{[\mathrm{Re}(E)-\delta\omega]^2+[\mathrm{Im}(E)]^2}.
\end{equation}
Here the band energy is 
\begin{equation}\label{eq:Ek}
E(k)=\bra{k}\mathbf{H}-i\gamma \ket{k},
\end{equation}
where $\gamma$ denotes all sources of loss in the absence of modulation.
For a given $k$, $I$ is a Lorentzian function of $\delta \omega$. Then, at each $k$, it is possible to deduce the values of $\mathrm{Re}(E)$ and $|\mathrm{Im}(E)|$ by fitting $I(\delta \omega)$ to the form of Eq.~\eqref{eq:lorentzian}.
In our setup, we use a cavity with a sufficiently large loss, i.e. a sufficiently large positive $\gamma$,
such that $\mathrm{Im}(E)<0$ is ensured for all $k$. For a $\gamma$ that is independent of $k$, $E(k)$ differs from the spectrum of $\mathbf{H}$ by a translation in the complex energy plane. The topological properties are not affected by such a translation. 

\begin{figure*}[t]
    \centering
    \includegraphics[width=1\textwidth]{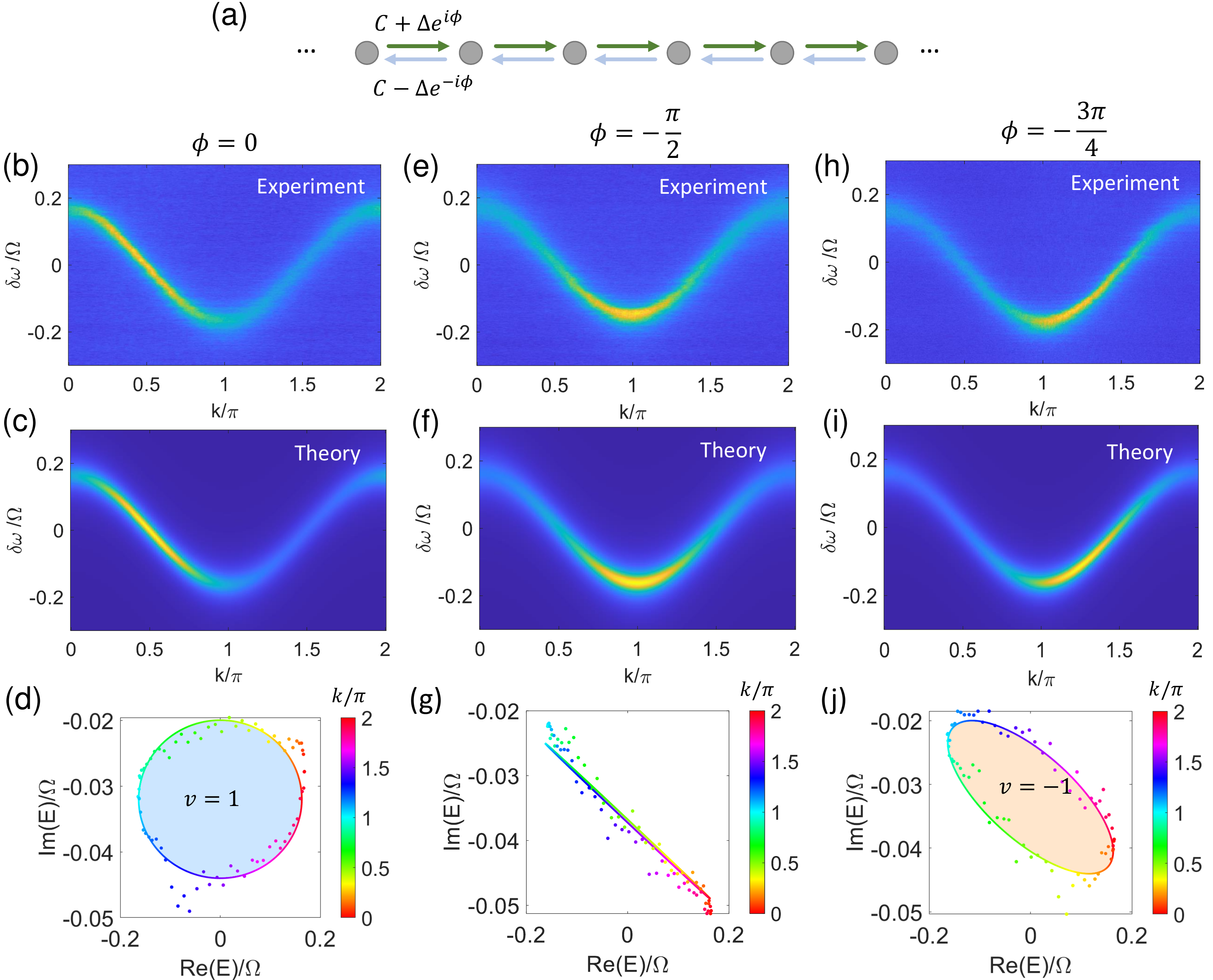}
    \caption{{\bf Experimental realization of a generalized Hatano-Nelson model and the measured topological windings.} (a)~Lattice model with nearest-neighbor non-Hermitian couplings. (b,e,h) Experimentally measured spectroscopy image for $\phi=0,\pi/2,3\pi/4$, respectively. (c,f,i) Theoretical predictions for the measurements shown in (b,e,h), respectively. (d,g,j)~The dots represent the band energy extracted from (b,e,h), respectively. The lines denote theoretical predictions. }
    \label{fig3}
\end{figure*}

To illustrate the specific process of the band-structure measurement, we show a representative experiment forming a lattice in synthetic frequency space as sketched in Fig.~\ref{fig2}(c). There is a 1st-order Hermitian coupling $\kappa_{\pm 1}=C$ and a 2nd-order anti-Hermitian coupling $\kappa_{\pm 2}=\pm \Delta \exp{(\pm i\phi)}$. By scanning $k$ over one Brillouin zone and $\delta \omega$ over one FSR $\Omega$, we obtain the $I(k,\delta \omega)$ readouts as shown by a 2D color plot in  Fig.~\ref{fig2}(d). Here we show two representative vertical slices at $k=\pi/4$ and $k=3\pi/4$ and plot the corresponding $I(\delta \omega)$ data in Figs.~\ref{fig2}(e,f), respectively. According to Eq.~\eqref{eq:lorentzian}, $\mathrm{Re}(E)$ and $\mathrm{Im}(E)$ can be obtained by a Lorentzian fit for each $k$-slice of $I$ [purple curves in Figs.~\ref{fig2}(e, f)].
We plot $\mathrm{Re}(E)$ and $\mathrm{Im}(E)$ thus obtained as the dots in Fig.~\ref{fig2}(g). The red curves in Fig.~\ref{fig2}(g) indicate the theoretically predicted form of the dispersion. The experimental results agree excellently with the theoretical predictions. Finally, we plot $E$ in the complex plane and visualize the winding [Fig.~\ref{fig2}(h)], where the experimental data points and the theoretical fitting curve are both color-coded to indicate the values of $k$. For the specific dispersion in this example, the complex band winds into a bow-tie shape. For $k$ going from $0$ to $2\pi$, the left-side loop (light orange shading) is traversed clockwise once with the winding number $v=-1$, and the right-side counter-clockwise loop (light blue shading) has $v=1$.

\begin{figure*}[t]
    \centering
    \includegraphics[width=1\textwidth]{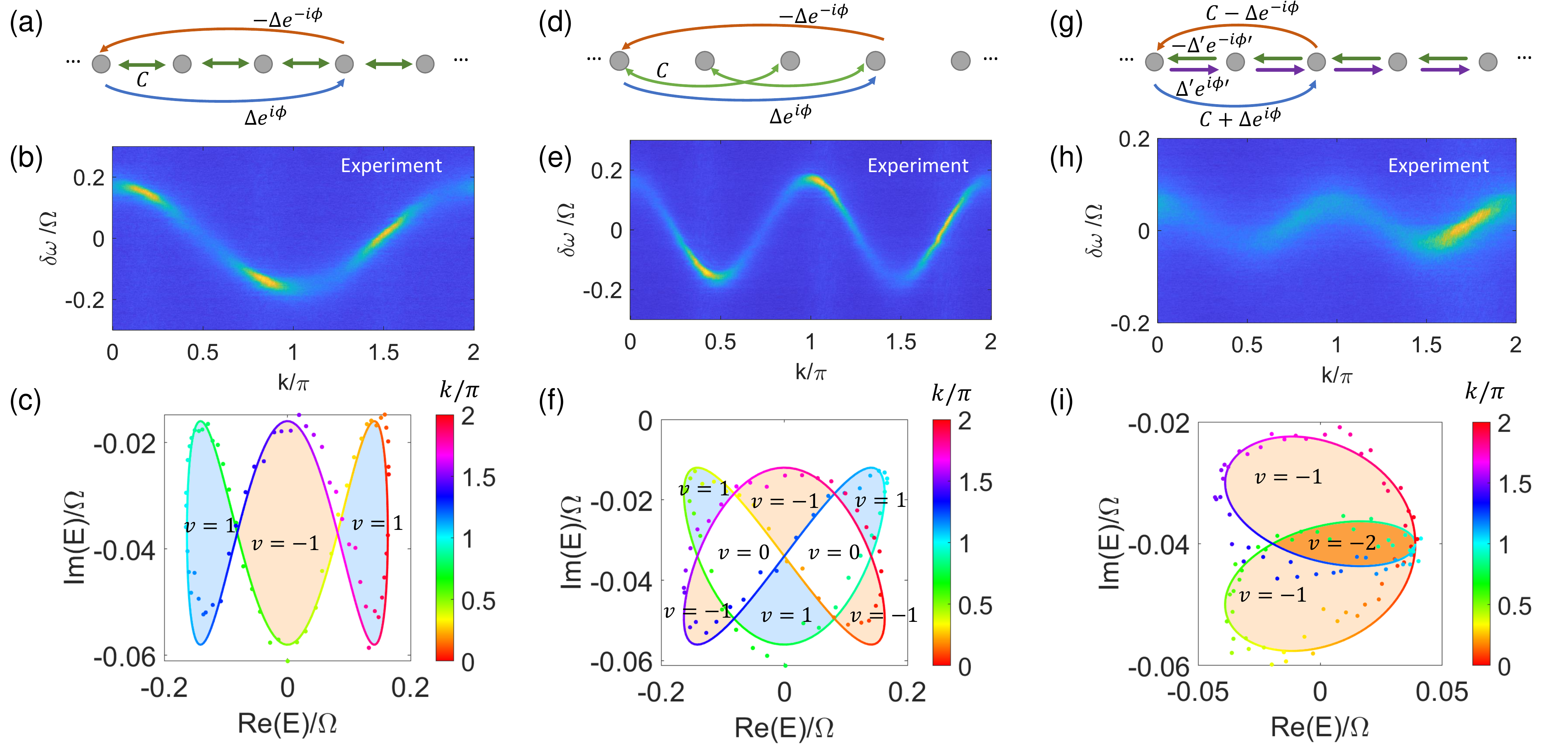}
    \caption{{\bf Realization and band winding measurements of non-Hermitian lattices incorporating long-range interactions.} (a) A lattice with 1st-order Hermitian coupling and 3rd-order anti-Hermitian coupling. (d) A lattice with 2nd-order Hermitian coupling and 3rd-order anti-Hermitian coupling. (g)~A lattice with 1st-order anti-Hermitian couplings and 2nd-order general non-Hermitian couplings. (b,e,h) Measured $I(k,\delta\omega)$ for the lattices in (a,d,g), respectively. (c,f,i) Deduced complex $E$ in complex plane for the lattices in (a,d,g), respectively. The dots are from experiments and the lines are from theory. }
    \label{fig4}
\end{figure*}

Strictly speaking, band structures can only be defined for \emph{infinite} systems that fulfill Bloch's theorem, whereas all practical structures in experiments are finite. 
Thus, in seeking to measure experimentally a band structure, one is always attempting to infer the property of an infinite system through the measurement on a corresponding finite system. For the Hermitian system this does not present an issue. Within the energy range of the band in an infinite system, the eigenstates of the corresponding finite system are also extended and provide an increasingly better approximation of the eigenstates of the infinite system as the size of the finite system increases. For the non-Hermitian system, on the other hand, the situation is more subtle. The eigenstates of the finite system can be drastically different from those of the infinite system.
As an extreme example, a finite non-Hermitian lattice truncated with an open boundary condition (OBC) can exhibit the non-Hermitian skin effect with all eigenstates localized on the edge, and the eigenspectra are thus drastically different from those of the corresponding infinite system~~\cite{Okuma2020,Helbig2020}.
Thus, it was noted that the winding number defined in Eq.~\eqref{eq:wn} can not be simply identified from the eigenspectrum of a finite non-Hermitian system~\cite{Gong2018}. And hence it may appear surprising that through Eq.~\eqref{eq:lorentzian} we are able to measure the band structure of an infinite system through the experiments on a finite system.  We emphasize that our experiments probe $E(k)$ as defined by Eq.~\eqref{eq:Ek}. $E(k)$ is \emph{not} an eigenenergy of the finite system, but rather the expectation value of the Hamiltonian of the finite system at state $\ket{k}$. The state $\ket{k}$ is always extended and moreover naturally approaches the Bloch state of an infinite system as the system size increases. Consequently, our experiments can provide a measurement of the band structure of an infinite system satisfying Bloch's theorem. As an illustration, in the supplementary Sec.~S1B and Fig.~S1 we consider a numerical experiment where we apply our technique for the extreme case as mentioned above where a finite non-Hermitian lattice is truncated by the OBC. Even though such a finite lattice exhibits a non-Hermitian skin effect, our technique still provides a faithful measurement of the band structure of the infinite system.

We now present an example where the winding of the band can be straightforwardly controlled.
In Fig.~\ref{fig3}(a) we sketch a lattice with nearest-neighbor non-Hermitian couplings in the form $\kappa_{\pm 1}=C\pm \Delta \exp{(\pm i\phi)}$. Importantly, the topological winding is largely determined by the value of the phase $\phi$, which can readily be tuned in our experiments by varying the relative phase between the amplitude and phase modulations. In Figs.~\ref{fig3}(b,c) we show the band structure measurement for $\phi=0$ accompanied by a theoretical prediction. This case serves as a faithful realization of the Hatano-Nelson model~\cite{Hatano1996}. The complex band encircles an elliptical area in the counterclockwise direction [Fig.~\ref{fig3}(d)] corresponding to a winding number $v=1$ (light blue shading). By contrast, if we have $\phi=-\pi/2$ [Figs.~\ref{fig3}(e,f)], the topology is trivial - the band forms a line in the complex plane [Fig.~\ref{fig3}(g)]. As we further choose $\phi=-3\pi/4$ [Figs.~\ref{fig3}(h,i)], the band becomes a tilted ellipse with a clockwise handedness, as shown in Fig.~\ref{fig3}(j). The winding number with respect to a reference energy inside the loop becomes $v=-1$.

A significant strength of synthetic dimensions lies in the ability to readily incorporate long-range couplings which can also be non-Hermitian. We harness this ability here to produce more complex windings. 
In Fig.~\ref{fig4}(a) we sketch a lattice model with a Hermitian 1st-order coupling $\kappa_{\pm 1}=C$, and a skew-Hermitian 3rd-order coupling  $\kappa_{\pm 3}=\pm \Delta \exp{(\pm i\phi)}$. This lattice has a dispersion $E(k)=2 C \cos(k)+2 i \Delta \sin (3k+\phi)-i\gamma$. Figs.~\ref{fig4}(b) and (c) show the measurement results for $\phi=0$. The extracted complex band winding is shown in Fig.~\ref{fig4}(c), which encloses three areas with the winding number $v=1,-1,1$ from left to right, respectively. 
As another example, in Fig.~\ref{fig4}(d) the Hermitian coupling is varied to 2nd-order compared to the lattice in Fig.~\ref{fig4}(a), and hence the dispersion becomes $E(k)=2 C \cos(2k)+2 i \Delta \sin (3k+\phi)-i\gamma$. We show a measurement and the extracted complex band winding with $\phi=-3\pi/4$ in Figs.~\ref{fig4}(e,f), respectively. The winding loop encloses three areas with $v=1$ (light blue shaded), three with $v=-1$ (light orange shading) and two areas with $v=0$ (no shading). These examples act as longer-range generalizations of the instances shown Fig.~\ref{fig3}. Additionally, since the winding number $v$ theoretically can take any integer value, to showcase the capability of implementing and measuring windings with $|v|>1$, we also present an instance as shown in Figs.~\ref{fig4}(g--i). The lattice model illustrated in Fig.~\ref{fig4}(g) incorporates non-Hermitian couplings of both 1st and 2nd orders, where we have $\kappa_{\pm 1}=\pm \Delta' \exp{(\pm i\phi')}$, $\kappa_{\pm 2}=C\pm \Delta \exp{(\pm i\phi)}$. The Bloch band is given by $E(k)=2C\cos(2k)+2i\Delta' \sin (k+\phi')+2i\Delta \sin (2k+\phi)-i\gamma$. In Fig.~\ref{fig4}(h) we show the measurement for the case $\phi=\phi'=\pi$, and the winding loop encloses not only areas with $v=-1$ (light orange shading) but also an area that winds twice clockwise, which corresponds to $v=-2$ (dark-orange shaded).
More example experimental results can be found in the Sec.~S3 of the supplementary.

In conclusion, we report the experimental observation of topological windings of the complex band energy in non-Hermitian lattices.
We show that non-Hermitian lattice Hamiltonians in the frequency synthetic space can be implemented through simultaneous amplitude and phase modulations of a ring resonator supporting multiple resonant modes. 
Our realization of non-Hermitian lattice Hamiltonians in synthetic space brings unprecedented flexibility to arbitrarily tailor the complex band structure fully combining non-Hermitian, long-range and complex-valued couplings. 
Future work could consider generalizations of such complex band winding to higher dimensions, using both real and synthetic space, or to multi-band lattices, such as those involving exceptional points~\cite{shen_topological_2018}.

\begin{acknowledgments}
We gratefully acknowledge David A.B. Miller for providing lab space and equipment and Meir Orenstein for useful discussions. This work is supported by a MURI project from the U. S. Air Force Office of Scientific Research (Grant No. FA9550-18-1-0379), and by a Vannevar Bush Faculty Fellowship from the U. S. Department of Defense (Grant No.  N00014-17-1-3030). 
\end{acknowledgments}


%

\newpage
\begin{center}
{\large \bf Supplementary Information}
\end{center}
\setcounter{equation}{0}
\setcounter{figure}{0}

\renewcommand{\theequation}{S\arabic{equation}}
\renewcommand{\thesection}{S\arabic{section}}
\renewcommand{\thefigure}{S\arabic{figure}}
\renewcommand{\thetable}{S\arabic{table}}
\renewcommand{\bibnumfmt}[1]{[#1]}
\renewcommand{\citenumfont}[1]{#1}


\section{Details on working principles}\label{sec:setup}

\subsection{Arbitrary one-band lattice implementation in synthetic space}\label{subsec:mod}

The implementation of Hermitian Hamiltonians in synthetic frequency space has been established in previous theoretical~\cite{Yuan2016, Ozawa2016,Yuan2018a,Dutt2020a,buddhiraju_arbitrary_2020} and experimental~\cite{Bell2017,Qin2018,Dutt2019, Dutt2020,Wang2020,Hu2020,Joshi2020,Li2021} works, by incorporating a phase modulation. It was also proposed that skew-Hermitian couplings can be formed in the frequency synthetic dimension via amplitude modulation~\cite{Yuan2018b}. In this section, we describe how the simultaneous use of phase and amplitude modulation driven by an electrical signal can facilitate arbitrary non-Hermitian couplings.

We consider a ring resonator with light propagating inside the resonator along a certain direction. We use $\psi$ to denote the electric field amplitude,  
and $t_R=L/v_g$ to denote the round-trip time, where $v_g$ is the group velocity of light in the ring of length $L$. Hence we have a set of discrete resonance modes for such a static resonator in the absence of dispersion, $\omega_n=\omega_0+n\Omega$, where $\Omega=2\pi/t_R$ is the free spectral range (FSR), $n$ is an integer indexing the modes and $\omega_0$ is the central frequency.

Now consider there exists a modulator inside the resonator that can perform both phase and amplitude modulation. There are two important assumptions we make here: (i) The modulator is spatially compact such that its length is negligible compared to the length $L$ of the ring. Thus we can describe the modulation by multiplying a $t$-dependent transmission factor $T(t)$ to the propagating light field passing the modulator. 
(ii) The periodic modulation signal is $t_R$-periodic in time, i.e.
\begin{equation}\label{eq:T_tR}
    T(t+t_R)=T(t).
\end{equation}
Next we check the single-frequency response of such a resonator system with an input port. We consider light input $\psi_{in}(\omega)$ at the input port with the frequency $\omega$. For convenience, we set the port to be placed right before the modulator, with an input/output coupling rate $\gamma_{in}=-\ln{(1-\eta)}/t_R$ for the in-coupling beam splitter with a splitting ratio $\eta/(1-\eta)$ in power ($\eta$ goes from the input port into the cavity). We aim to establish the relation
\begin{equation}\label{eq:in_cav_relation}
    \psi=g \sqrt{\eta}\,\psi_{in}(\omega)=g \sqrt{1-e^{-\gamma_{in}t_R}} \psi_{in}(\omega),
\end{equation}
where $\psi$ is the intracavity steady-state electric field amplitude at the same position as the input port. Here $g$ can be calculated by summing up the field amplitude of light having traveled different numbers of round-trips:
\begin{equation}\label{eq:gs}
\begin{split}
    g&=1+T(t) e^{-i\omega t_R-\gamma_0 t_R-\frac{\gamma_{in}t_R}{2}}+T^2(t)  e^{2(-i\omega t_R-\gamma_0 t_R-\frac{\gamma_{in}t_R}{2})}+T^3(t)  e^{3(-i\omega t_R-\gamma_0 t_R-\frac{\gamma_{in}t_R}{2})}+\dots\\
    &=\sum_{\tau=0}^\infty \left[T(t) e^{-i\omega t_R-\gamma_0 t_R-\frac{\gamma_{in}t_R}{2}}\right]^\tau,
\end{split}
\end{equation}
where $\tau$ is a slow-time variable counting how many times light passes the modulator. Here we have utilized the $t_R$-periodicity of $T(t)$ as given in Eq.~\eqref{eq:T_tR}. Eq.~\eqref{eq:gs} forms a geometric series; if it converges (i.e. the system is lossy at all times $t$), we have
\begin{equation}\label{eq:g_exact}
    g=\frac{1}{1-\left[T(t) e^{-i\omega t_R-\gamma_0 t_R-\frac{\gamma_{in}t_R}{2}}\right]}.
\end{equation}
Note that $g$ is essentially a frequency-domain Green's function for a system evolving with the discrete time $\tau$, but it has a periodic $t$-dependency passed on from $T(t)$, $g(t+t_R)=g(t)$. Eq.~\eqref{eq:gs} can also be regarded as a discrete-time Fourier transform or a $z$-transform.

Now we consider the specific modulation waveforms that the modulator generates. 
We assume that the modulator produces simultaneous phase-modulation $T_{Ph}$ and amplitude-modulation $T_{Am}$, which together constitute the modulator transmission
\begin{equation}\label{eq:T}
    T=T_{Am} T_{Ph}.
\end{equation}
Then, we discuss the specific forms of phase and amplitude modulations, respectively. The phase modulation can be modeled by
\begin{equation}\label{eq:Tph}
T_{Ph}= e^{-i \delta \Phi (t)}=e^{-i \sum_m A_m \cos (m \Omega t+\alpha_m)}, 
\end{equation}
where $\delta \Phi(t)=\sum_m A_m \cos (m \Omega t+\alpha_m)$ with $m$ being integers. 
On the other hand, the amplitude modulation has the form
\begin{equation}\label{eq:Tam_orig}
T_{Am}=1+\sum_m B_m \sin (m \Omega t+\beta_m). 
\end{equation}
We note that in Eq.~\eqref{eq:Tam_orig} $T_{Am}$ exceeds unity at some time intervals. The form of Eq.~\eqref{eq:Tam_orig} in practice can be achieved by using an amplitude modulator where the loss is modulated periodically, together with an amplifier that provides a constant gain over time. 
Here, since we work in a regime with $|B_m| \ll 1$, we can express $T_{Am}$ in the following form
\begin{equation}\label{eq:Tam}
    T_{Am}=e^{\ln{[1+\sum_m B_m \sin (m \Omega t+\beta_m)]}}\approx e^{\sum_m B_m \sin (m \Omega t+\beta_m)},
\end{equation}
where we take the first-order approximation in the Taylor expansion. 
Then, by inserting Eq.~\eqref{eq:Tph} and Eq.~\eqref{eq:Tam} into Eq.~\eqref{eq:T}, we find
\begin{equation}\label{eq:Tfinal}
    T = e^{-i \sum_m A_m \cos (m \Omega t+\alpha_m)+\sum_m B_m \sin (m \Omega t+\beta_m)}.
\end{equation}

Now, we insert the transmission of the modulator described by Eq.~\eqref{eq:Tfinal} into the expression of Green's function $g$ given in Eq.~\eqref{eq:g_exact}, obtaining
\begin{equation}\label{eq:g}
    g=\frac{1}{1-e^{-i \sum_m A_m \cos (m \Omega t+\alpha_m)+\sum_m B_m \sin (m \Omega t+\beta_m)  -i\omega t_R-\gamma_0 t_R-\frac{\gamma_{in}t_R}{2}}}.
\end{equation}

In the paragraphs below we will derive the Hamiltonian using the Green's function in Eq.~\eqref{eq:g}. The discussion will be split into two cases: (a) we still treat $\tau$ as a discrete variable and use it as the evolution time index; (b) in the limit of small $\omega t_R$ we approximately consider the regime that $\tau$ is a continuous time variable. 

{\bf a.~Discrete-time analysis} 

In our experiments we operate in a regime where the variation of the field per trip around the ring is small, i.e. $|-i \sum_m A_m \cos (m \Omega t+\alpha_m)+\sum_m B_m \sin (m \Omega t+\beta_m)-\gamma_0 t_R-\gamma_{in}t_R/2| \ll 1$; thereby we can take the first-order approximation in the Taylor expansion of $\exp{[-i \sum_m A_m \cos (m \Omega t+\alpha_m)+\sum_m B_m \sin (m \Omega t+\beta_m)-\gamma_0 t_R-\gamma_{in}t_R/2]}$ in Eq.~\eqref{eq:g}. Then, we have 
\begin{equation}\label{eq:g_approx_d}
\begin{split}
    g&\approx g_d \\
    &= \frac{1}{1-e^{-i\omega t_R}-e^{-i\omega t_R}\left[-i \sum_m A_m \cos (m \Omega t+\alpha_m)+\sum_m B_m \sin (m \Omega t+\beta_m) -\gamma_0 t_R-\frac{\gamma_{in}t_R}{2}\right]}.
\end{split}
\end{equation}
We use $H_d$ to denote the Hamiltonian that governs the evolution of $\psi_{\tau}$ under the discrete time steps $\tau$. The frequency-domain discrete-time Schr\"odinger equation incorporating driving is given by
\begin{equation}\label{eq:schro_fd}
    i(e^{i\omega t_R}-1  ) \psi= \left(H_d-i\frac{\gamma_{in}t_R}{2} \right)  \psi +ie^{i\omega t_R} \sqrt{1-e^{-\gamma_{in}t_R}}\psi_{in},
\end{equation}
which is the discrete-time Fourier transform of the difference equation
\begin{equation}
    i(\psi_{\tau+1} - \psi_{\tau} ) = \left( H_d - {i\gamma_{in}t_R\over 2}\right) \psi_{\tau} + i \sqrt{1-e^{-\gamma_{in}t_R}} \psi_{in,\tau+1} .
\end{equation}
Here we can define a discrete-time forward difference operator $\Delta_\tau$ that fulfills $\Delta_\tau \psi_\tau=\psi_{\tau+1}-\psi_{\tau}$.
Then, inserting Eq.~\eqref{eq:in_cav_relation} and Eq.~\eqref{eq:g_approx_d} into Eq.~\eqref{eq:schro_fd} we find
\begin{equation}
    H_d=\sum_m A_m \cos (m \Omega t-\alpha_m)+i\sum_m B_m \sin (m \Omega t+\beta_m)-i\gamma_0 t_R.
\end{equation}
Without loss of generality, in the following analysis we take $\gamma_0=0$. 
Then, we can use the ansatz
\begin{equation}\label{eq:ans_d}
     \psi_\tau(t)= \sum_n a_{n,\tau} e^{i\omega_n  t}.
\end{equation}
Now consider the discrete-time $\tau$-dependent Schr\"odinger equation
\begin{equation}
    i\Delta_\tau {\psi_\tau}= H_d {\psi_\tau}.
\end{equation}
By inserting the ansatz in Eq.~\eqref{eq:ans_d} to such a Schr\"odinger equation, we find
 \begin{equation}
     i \Delta_\tau  a_{n,\tau}=\sum_m \left(\frac{A_m}{2} e^{i\alpha_m}+\frac{B_m}{2} e^{i\beta_m}\right) a_{n-m,\tau}+\sum_m \left( \frac{A_m}{2} e^{-i\alpha_m}-\frac{B_m}{2} e^{-i\beta_m} \right) a_{n+m,\tau}.
 \end{equation}
In the lattice space (frequency-mode space), this corresponds to a tight-binding model with general couplings between lattice sites.

{\bf b.~Continuous-time regime}

If we take the first-order Taylor expansion of the entire second term in the denominator of Eq.~\eqref{eq:g}, we can obtain
\begin{equation}\label{eq:g_approx}
    g\approx g_c= \frac{1}{i \sum_m A_m \cos (m \Omega t-\alpha_m)-\sum_m B_m \sin (m \Omega t+\beta_m)+i\omega t_R+\gamma_0 t_R+\frac{\gamma_{in}t_R}{2}}.
\end{equation}
In doing so, we have assumed $|\omega t_R|$ is small. Equivalently, this means we consider a regime where the discrete slow time $\tau$ can be approximately treated as a continuous variable, and the following analysis will be in this regime. If we multiply both sides of Eq.~\eqref{eq:schro_fd} by $1/\sqrt{t_R}$, we have
\begin{equation}
    i\frac{e^{i\omega t_R}-1 }{t_R} \tilde{\psi}= \left(\frac{H_d}{t_R}-i\frac{\gamma_{in}}{2} \right)  \tilde{\psi} +i e^{i\omega t_R} \sqrt{\frac{1-e^{-\gamma_{in}t_R}}{t_R}}\psi_{in}.
\end{equation}
Here we have defined $\tilde{\psi}=\psi \sqrt{t_R}$. This is because, in the continuous-time limit, light in the resonator $|\tilde{\psi}|^2$ becomes an energy flux, whereas the input light $|\psi_{in}|^2$ is still measured in intensity since it remains a propagating wave.
With the small $t_R$, we also have $(e^{i\omega t_R}-1 )/t_R \approx i \omega$ and $e^{i\omega t_R} \sqrt{(1-e^{-\gamma_{in}t_R})/t_R} \approx \sqrt{\gamma_{in}}$. Hence we end up with the following frequency-domain Schr\"odinger equation including an external driving term,
\begin{equation}\label{eq:schro_f}
    -\omega  \tilde{\psi}= \left( H_c  -i\frac{\gamma_{in}}{2} \right)\tilde{\psi} +i\sqrt{\gamma_{in}}\psi_{in},
\end{equation}
where the Hamiltonian is given by
\begin{equation}
    H_c=\frac{H_d}{t_R}=\sum_m \frac{A_m}{t_R} \cos (m \Omega t+\alpha_m)+i\sum_m \frac{B_m}{t_R} \sin (m \Omega t+\beta_m)-i\gamma_0.
\end{equation}
Without loss of generality, in the following analysis we take $\gamma_0=0$. We note that Eq.~\eqref{eq:schro_f} is the Fourier transform of the time-domain Schr\"odinger equation with driving,
\begin{equation}
   i\frac{\partial}{\partial \tau} \tilde{\psi}(\tau)= \left( H_c  -i\frac{\gamma_{in}}{2} \right)\tilde{\psi}(\tau) +i\sqrt{\gamma_{in}}\psi_{in}.    
\end{equation}
Then, we can also include the periodic $t$-dependency of $\tilde{\psi}$ in steady state with the ansatz
\begin{equation}\label{eq:psi_expan}
     \tilde{\psi}(\tau,t)= \sum_n a_n(\tau) e^{i\omega_n  t}.
\end{equation}
By inserting this ansatz into the $\tau$-dependent Schr\"odinger equation
\begin{equation}
    i\frac{\partial}{\partial \tau} \tilde{\psi}= H_c \tilde{\psi},
\end{equation}
we end up with a set of equations
 \begin{equation}\label{eq:eom}
     i \frac{\mathrm{d}}{\mathrm{d} \tau} a_n=\sum_m ( C_{m} e^{i\alpha_m}+\Delta_{m} e^{i\beta_m}) a_{n-m}+\sum_m ( C_{m} e^{-i\alpha_m}-\Delta_{m} e^{-i\beta_m} ) a_{n+m},
 \end{equation}
where $C_m=A_m/2 t_R$ and $\Delta_m=B_m/2 t_R$. Eq.~\eqref{eq:eom} essentially resembles the equations of motions in such a continuous-time regime. 
Then we can see that the frequency-mode space Hamiltonian is 
\begin{equation}~\label{eq:hamil}
    \mathbf{H}=\sum_{m,n} \kappa_{+m}\mathbf{a_{n+m}^\dagger}\mathbf{a_{n}}
    +  \kappa_{-m}\mathbf{a_{n}^\dagger}\mathbf{a_{n+m}},
\end{equation}
where $\kappa_{\pm m}=C_{m} e^{i\alpha_m} \pm \Delta_{m} e^{\pm i\beta_m}$, and $\mathbf{a_{n}^\dagger}$ ($\mathbf{a_{n}}$) is the creation (annihilation) operator of the $n$-th frequency mode. Therefore, we have established that the discrete frequency modes in such a frequency- and amplitude- modulated resonator serve as an implementation of a tight-binding lattice with the Hamiltonian given in Eq.~\eqref{eq:hamil}. 

\subsection{Non-Hermitian band-structure measurement via expectation values}\label{subsec:bs}

Here we further elaborate on the band-structure measurement approach that is suitable for finite non-Hermitian lattices. As mentioned in the main text, an attempt to experimentally measure a band structure necessarily needs to infer the property of an infinite system from a finite system. If the finite lattice is under the periodic boundary condition (PBC), it is well known that the eigenvalues do faithfully resemble that of the infinite counterpart. However, the cases with other boundary conditions can be trickier. In particular, for a finite non-Hermitian lattice that is not under strict PBC, the eigenstates are generally not extended. This presents a challenge in the band structure measurements of non-Hermitian lattices.

The approach we take here is based on the expectation value $E(k)$ obtained via projecting the non-Hermitian Hamiltonian to $\ket{k}$ prepared in the excitation and measurement process, where the Bloch functions $\ket{k}$ are always extended regardless of how the eigenstates behave. This enables a measurement of the band structure using a finite lattice without the PBC. In the aspect of the general measurement scheme,
complex expectation-value measurements were utilized previously in determining high-dimensional quantum states~\cite{Bolduc2016}, where the measured object (e.g. density matrix) is Hermitian while the measurement operator is non-Hermitian. Here, by contrast, the measured system is non-Hermitian and the projective measurements with the Hermitian projector $\ket{k}\bra{k}$ can yield complex expectation-value outcomes. 
In the context of non-Hermitian topology characterization,
our approach is in contrast with eigenspectrum measurements under strict PBC~\cite{Helbig2020}, which has no $k$-resolution and thus cannot characterize the winding number unless the eigenstate associated with each eigenvalue is fully determined.
Our scheme is related to but more direct than an earlier theoretical proposal to infer the non-Hemitian band winding~\cite{Gong2018} via Bloch oscillations~\cite{Wimmer2015}.

We now establish our approach firstly in a general form. Consider a lattice Hamiltonian $\mathbf{H}$ with $N$ sites that can be constructed by subtracting a matrix $\mathbf{M}$ from the PBC Hamiltonian $\mathbf{H^{PBC}}$, with $\mathbf{M}$ describing the deviation in boundary condition,
\begin{equation}\label{eq:pbc2general}
\mathbf{H}=\mathbf{H^{PBC}}-\mathbf{M}.     
\end{equation}
Then, the expectation value is what we aim to measure:
\begin{equation}\label{eq:E_n}
E_n=\bra{k_n}\mathbf{H}\ket{k_n}=\lambda_n^{\mathrm{PBC}}-\bra{k_n}\mathbf{M}\ket{k_n},    
\end{equation}
where $\lambda_n^{\mathrm{PBC}}$ denotes the PBC eigenspectra that are known to faithfully reflect the band structure. It is reasonable to assume that the boundary condition is only added to a finite number of elements in the Hamiltonian, i.e. there are a finite, $N$-independent number of non-zero elements in $\mathbf{M}$, thereby $\bra{k_n}\mathbf{M}\ket{k_n}\propto 1/N$. Hence we can immediately see that 
\begin{equation}
\lim_{N\to \infty} E_n= \lambda_n^{\mathrm{PBC}},
\end{equation}
where the PBC eigenspectrum in this limit is precisely the band structure.
In practice, for sufficiently large $N$, the approximation $E_n \approx \lambda_n^{\mathrm{PBC}}$ can be made.

The lattice we implement in synthetic frequency space has no sharp boundary for over $10^3$ lattice sites (see Sec. S2 for details), yet we stress that this is not a necessary condition to perform such a band-structure measurement. In fact, as we mentioned in the main text, even if the lattice is under open boundary condition (OBC) and thus can exhibit the so-called non-Hermitian skin effect~\cite{Lee2016,Yao2018,Longhi2019a,Longhi2019,Borgnia2020,Okuma2020}, the expectation-value measurement can still yield the band structure defined for the infinite counterpart that fulfills Bloch's theorem. 

Here we analytically show and numerically verify this specific argument on OBC. Consider a PBC lattice with $N$ sites and up to $M$ orders of coupling. Since in this case the discrete Bloch functions $\ket{k_n}$ are its eigenvectors, we can expand its Hamiltonian $\mathbf{H^{PBC}}$ into an eigen-basis by
\begin{equation}
    \mathbf{H^{PBC}}=\sum_{n=1}^N \lambda^{\mathrm{PBC}}_n \ket{k_n}\bra{k_n},
\end{equation}
where $\lambda^{\mathrm{PBC}}_n$ are the eigenvalues given by
\begin{equation}
   \lambda^{\mathrm{PBC}}_n= 2\sum_{m=1}^M C_m \cos (m  k_n+\alpha_m)+2i\sum_{m=1}^M \Delta_m \sin (m k_n+\beta_m).
\end{equation}
Such an eigenspectrum can be obtained from the general 1D lattice Hamiltonian given in Eq.~\eqref{eq:hamil} under PBC.
Now we construct a lattice with an OBC based on such a Hamiltonian with a PBC, using the formalism as described in Eq.~\eqref{eq:pbc2general}, by subtracting a matrix $\mathbf{M}$ from $\mathbf{H^{PBC}}$,
\begin{equation}\label{eq:HM}
    \mathbf{H^{OBC}}=\mathbf{H^{PBC}}-\mathbf{M},
\end{equation}
where
\begin{equation}\label{eq:M}
    \mathbf{M}=\sum_{m=1}^M \sum_{q=0}^{m-1} \left( \kappa_{-m}\ket{l-q}\bra{l-q+m}+\kappa_{+m}\ket{l-q+m}\bra{l-q} \right).
\end{equation}
The matrix $\mathbf{M}$ describes the removal of couplings when the PBC lattice is split between the $l$-th site and the $(l+1)$-th site, forming an OBC lattice.
From Eq.~\eqref{eq:E_n},
it can be seen that the difference between the expectation value $E_n$ and the PBC eigenvalue $\lambda^{\mathrm{PBC}}_n$ is fully captured by $\bra{k_n}\mathbf{M}\ket{k_n}$, which can be calculated by inserting Eq.~\eqref{eq:M},
\begin{equation}\label{eq:kMk}
\bra{k_n}\mathbf{M}\ket{k_n}=\frac{1}{N}\sum_{m=1}^M m (\kappa_{-m} e^{-i m k_n}+\kappa_{+m} e^{i m k_n}).
\end{equation}
As can be seen from Eq.~\eqref{eq:kMk}, for a given lattice with a set of coupling parameters, $\bra{k_n}\mathbf{M}\ket{k_n}$ scales with $1/N$ for such an OBC case; therefore the deviation between $E_n$ and $\lambda^{\mathrm{PBC}}_n$ indeed becomes negligible if $N$ is sufficiently large. 

\begin{figure}[t]
    \centering
    \includegraphics[width=0.9\columnwidth]{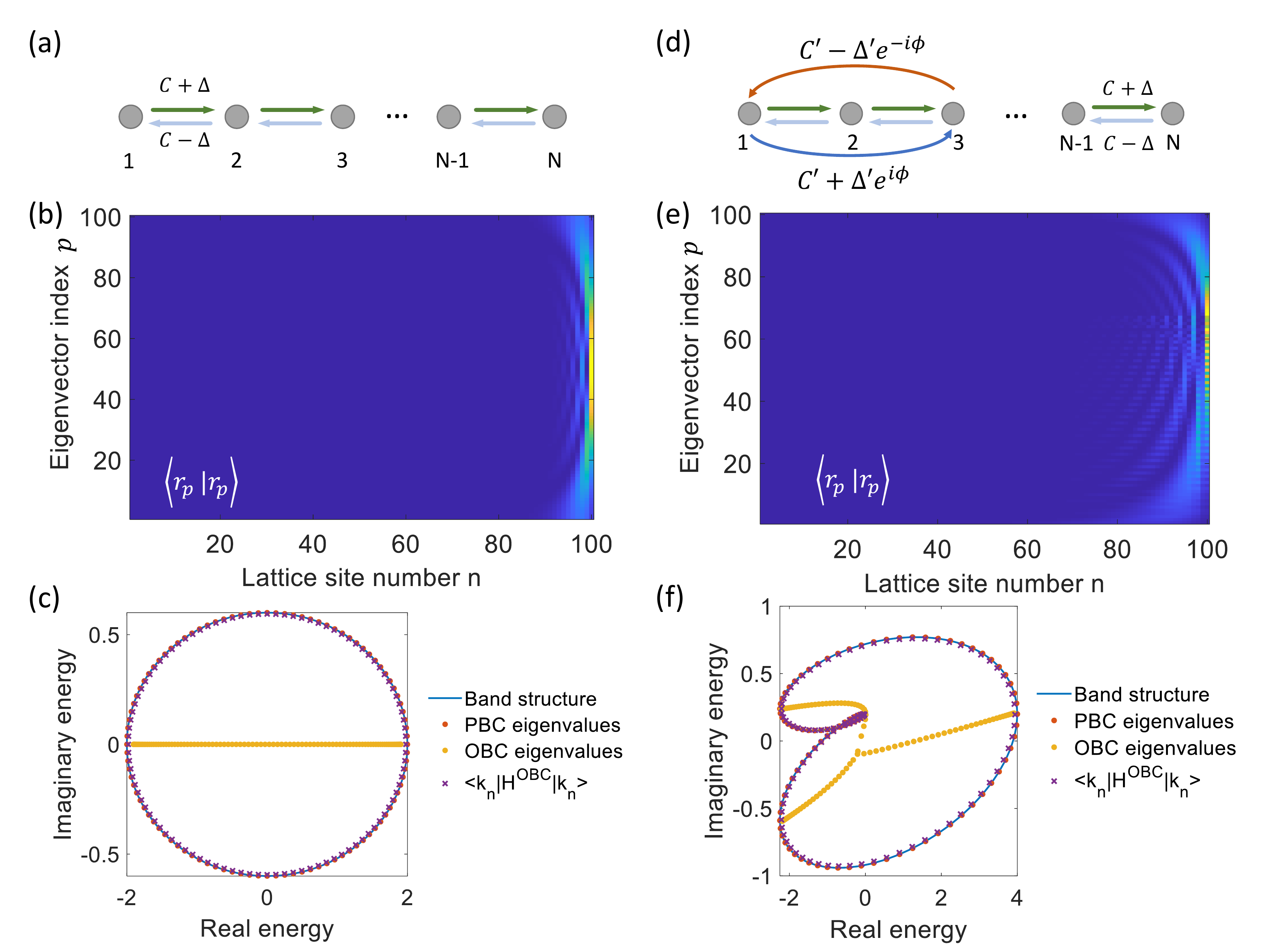}
    \caption{Numerical examples of expectation-value band-structure measurement in OBC non-Hermitian lattices. (a,d) Lattice model sketch. (b,e) Right-eigenvector probability in lattice-site space for the models in (a,d), respectively. (c,f) Band structure, PBC eigenvalues, OBC eigenvalues and the expectation value $E_n=\bra{k_n}\mathbf{H^{OBC}}\ket{k_n}$ for the models in (a,d), respectively.}
    \label{fig:eig}
\end{figure}

To verify and intuitively illustrate this conclusion on OBC lattices, here we provide numerical examples. We base the numerical experiments on a lattice Hamiltonian with $N=100$. In the first example, we consider the Hatano-Nelson model as sketched in Fig.~\ref{fig:eig}(a), which has 1st-order non-Hermitian couplings $\kappa_{\pm 1}=C\pm \Delta$ with $C=1$ and $\Delta=0.3$. In Fig.~\ref{fig:eig}(b) we plot $\braket{r_p|r_p}$ for the right-eigenvectors $\ket{r_p}$ of this lattice under OBC, where $p$ is an integer indexing the eigenvectors. We can see that all $\ket{r_p}$ are localized at the right-side edge, indicating the presence of the non-Hermitian skin effect. In Fig.~\ref{fig:eig}(c), in the complex energy plane, we plot the eigenvalues under PBC (red dots) and OBC (yellow dots). In the PBC case, the eigenvalues all exactly lie on the band structure (blue curve), winding along an ellipse and thus having nontrivial topology. By contrast, the OBC eigenvalues all stay on the real-energy axis, and thus do not exhibit non-trivial winding. Nevertheless, as we also plot the values of $E_n=\bra{k_n}\mathbf{H^{OBC}}\ket{k_n}$ (purple crosses), it can be seen that even for such a lattice in the extreme case of an OBC, the expectation value $E_n$ can still serve as a good approximation of the complex band structure. 

In another example shown in Figs.~\ref{fig:eig}(d--f), we keep the same 1st-order couplings yet add a 2nd-order coupling $\kappa_{\pm 2}=C'\pm \Delta'\exp(\pm i \phi)$ with $C'=1$, $\Delta'=0.2$ and $\phi=\pi/6$. The same conclusion still holds, that the expectation value $E_n(k)$ can indeed well represent the band structure even in an extreme case where the eigenstates are highly localized.


\subsection{Extracting complex energy from $k$-projections of Green's functions}

As discussed in the main text, the expectation value $E(k)=\bra{k}\mathbf{H}\ket{k}$ is determined in our experiments via projecting the Green's function $\mathbf{G}$ to $\ket{k}$, where the direct observable is $I(k,\delta \omega)=|\bra{k}\mathbf{G}\ket{k}|^2=|G(k,\delta \omega)|^2$. Here, if we take the continuous-time limit as discussed in Sec.~\ref{subsec:mod}, then $G (k,\delta\omega)\approx g_c (t=k/\Omega, \omega=\omega_n-\delta \omega)t_R$, where $g_c$ is given in Eq.~\eqref{eq:g_approx}.

Now we provide a more specific analysis on the condition to use the Green's function $g_c$ in Eq.~\eqref{eq:g_approx} to approximate $g$ in Eq.~\eqref{eq:g}. For convenience, we define
\begin{equation}
    E(k)= \sum_m \frac{A_m}{t_R} \cos (m k+\alpha_m)+i\sum_m \frac{B_m}{t_R} \sin (m k+\beta_m) -i\gamma_0 -i\frac{\gamma_{in}}{2},
\end{equation}
which is the object of measurement.
Then, we can rewrite Eq.~\eqref{eq:g} as
\begin{equation}
  g\left(t=\frac{k}{\Omega}, \omega=\omega_n-\delta \omega\right)=\frac{1}{1-e^{-i2\pi\{-\delta\omega + \mathrm{Re}[E(k)]\}/\Omega+2\pi\mathrm{Im}[E(k)]/\Omega}}, 
\end{equation}
where we have used $t_R=2\pi/\Omega$.
Note that the actual condition to treat the second term in the denominator perturbatively is (i) $|\delta\omega -\mathrm{Re}[E(k)]| \ll \Omega$ and (ii) $|\mathrm{Im}[E(k)]|\ll\Omega $. The condition (ii) is immediately fulfilled by looking at our experimental parameters since we have $\mathrm{Im}[E(k)] < 0.07 \Omega$ throughout the measurements. For condition (i), although we sweep a large range of $\delta \omega$, note that with the small $\mathrm{Im}[E(k)]$, our Lorentzian fitting is mostly relevant around a small frequency range centered at $\mathrm{Re}[E(k)]$. 
Therefore, we can end up using the form
\begin{equation}
  G(k, \delta{\omega} )=\frac{1}{i\{\mathrm{Re}[E(k)]-\delta\omega  \}-\mathrm{Im}[E(k)]},    
\end{equation}
which is a good approximation for $\delta \omega$ around the peak center $\mathrm{Re}(E)$ in the measured $I(k,\delta \omega)$.

\section{Details on experimental methods}\label{sec:setup}

\begin{figure}[t]
    \centering
    \includegraphics[width=0.85\columnwidth]{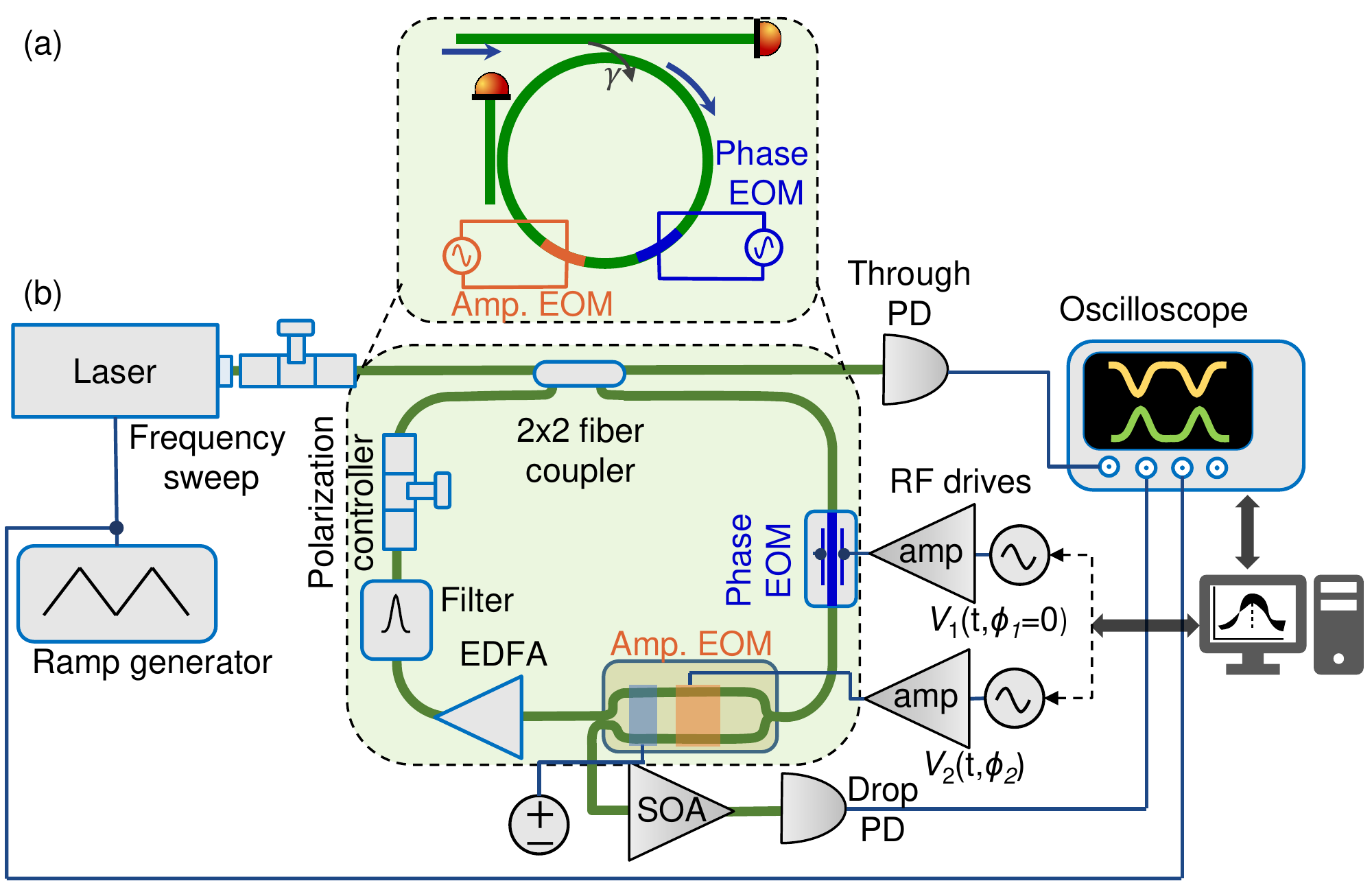}
    \caption{Schemetic of the experimental setup: (a) conceptual sketch, (b) detailed setup.}
    \label{fig:setup}
\end{figure}

In this section, we provide details on the experimental setup for this work. As mentioned in the main text and sketched conceptually in Fig.~\ref{fig:setup}(a), the experiment is done with an optical ring resonator undergoing both phase and amplitude modulations via electro-optic modulators (EOMs).

In Fig.~\ref{fig:setup}(b) we show a detailed layout of the setup we used. Specifically, the implementation of the optical ring resonators and the auxiliary waveguides are all based on optical fibers. The ring has an FSR of $\Omega=2\pi \times 5.335$ MHz. The ring is excited by a CW laser in the telecommunications C-band with a linewidth of 2.8 kHz and a center wavelength of 1542.057 nm. The laser frequency is swept by a function generator to obtain a ramp signal, where we generate a triangular signal with $V_{pp}=0.5$ V at a frequency of $160$ Hz. The fiber-coupled laser first goes through a polarization controller and then enters the fiber ring resonator by a $2\times 2$ fiber coupler (beam splitter). The transmitted light at this coupler is sent to a photodiode (namely `Through PD' in the figure) for monitoring. In the fiber ring resonator [green shaded in Fig.~\ref{fig:setup}(b)], we utilize the clockwise propagating resonance modes. The in-coupled light goes through phase and amplitude EOMs, followed by an Er-doped fiber amplifier (EDFA) to partially compensate for cavity round-trip loss, followed by a dense-wavelength division multiplexing (DWDM) band-pass filter (Channel 44, center wavelength 1542.14 nm, transmission bandwidth 26.5 GHz). The passband of this filter supports at least $4.9 \times 10^{3}$ frequency modes with the FSR used. Finally, a polarization controller inside the cavity ensures that the polarization orientation in the fiber is unchanged after one round-trip. The amplitude modulation inside the EOM is implemented by a lithium-niobate-waveguide Mach-Zehnder interferometer (MZI), incorporating a phase-modulation module and a DC bias, with the latter used for operating the amplitude modulator in the linear regime of the MZI fringe (we formulate this in a more concrete manner in the next paragraph). We make use of the extra output after the modulator as a drop port that samples a portion of the in-cavity light. Light from this port is guided through a semiconductor optical amplifier (SOA) and then detected by a 5-GHz-bandwidth photodiode (denoted as `Drop PD'). The detected signals are all monitored and captured by an oscilloscope that connects to a computer. The arbitrary electrical signals for driving the phase and amplitude modulations are generated via an instrument based on field-programmable gate arrays (FPGAs), and amplified by radio-frequency (RF) amplifiers.

{\bf MZI-based amplitude modulation.}
Here we provide some details on the amplitude modulator based on an MZI. As mentioned above, the amplitude modulation we use is done by an MZI incorporating a phase modulator and a DC bias. Without loss of generality, here we formulate it in the most commonly used push-pull configuration, i.e. the two arms of the MZI undergo the same phase modulation and bias, but with a flipped sign~\cite{haus_waves_1984}. Consider the phase modulation $\phi_a=\phi_0+\sum_m J_m \sin (m \Omega t+\beta_m)$ with the static phase $\phi_0$ provided by a DC bias voltage. One arm of the MZI is modulated with $\phi_a/2$ while that of the other arm with $-\phi_a/2$. We can check the transmission from either port of the MZI, e.g. 
\begin{equation}
    T_{MZI}
    =\sin \left(\frac{\phi_a}{2}\right).
\end{equation}
Here we have assumed the global phase factor is unity. In the absence of modulation, i.e. $J_m=0$, if we use $\phi_0=\pi/2$, it is easy to calculate that $|T_{MZI}|^2=1/2$. This means that half of the input power exits from this port. In the vicinity of this point, the response of the output amplitude to a phase change is approximately linear. If $|\sum_m J_m \sin (m \Omega t+\beta_m)|\ll 1$ we can take the first-order approximation in the Taylor expansion of $\sin (\phi_a/2)$ around $\phi_0=\pi/2$, , and obtain
\begin{equation}
    T_{MZI}\left(\phi_0=\frac{\pi}{2}\right)\approx \frac{1}{\sqrt{2}} + \frac{1}{2\sqrt{2}}\sum_m J_m \sin (m \Omega t+\beta_m).
\end{equation}
If there is gain after the MZI-modulator with the amplification factor $\zeta=\sqrt{2}$, then we can obtain the desired amplitude modulation form
\begin{equation}
    T_{Am}=\zeta T_{MZI}(\phi_0=\frac{\pi}{2})= 1 + \sum_m \frac{J_m}{2} \sin (m \Omega t+\beta_m).
\end{equation}
Then the amplitude modulation is exactly the same as that used in Eq.~\eqref{eq:Tam_orig}, with $B_m=J_m/2$.

{\bf Laser frequency sweeping and band structure spectroscopy.}
As mentioned in the main text, the measurement is based on a $k$- and detuning-resolved intensity measurement from a drop port of the ring resonator, $I(k,\delta \omega)$. While $k= t\Omega$ is essentially time, we can also make the detuning $\delta \omega$ linearly time-dependent to achieve the both $k$- and $\delta \omega$-resolved measurement with a single-shot multiplexed temporal detection. This can be done by sweeping the frequency of the input laser light such that $\delta \omega (t)= \xi t+ \omega_c$ where $\xi$ represents the slope and $\omega_c$ is a constant frequency. Then the measured time-dependent $I(t)$ is both $\delta \omega$- and $k$-encoded~\cite{Dutt2019}. 
When processing the signal $I(t)$, we can sequentially take out each $t\in [2\pi n/\Omega, 2\pi (n+1)/\Omega)$ time interval (corresponding to $k\in [0,2\pi)$) and assign $\delta \omega \approx \xi 2\pi (n+1/2)/\Omega + \omega_c$. Then we can obtain the $I(\delta \omega,k)$ data. 

\section{Supplementary experimental results}\label{sec:exp}

In this section, we present more experimental examples to supplement the main manuscript.

\begin{figure}[t]
    \centering
    \includegraphics[width=0.9\columnwidth]{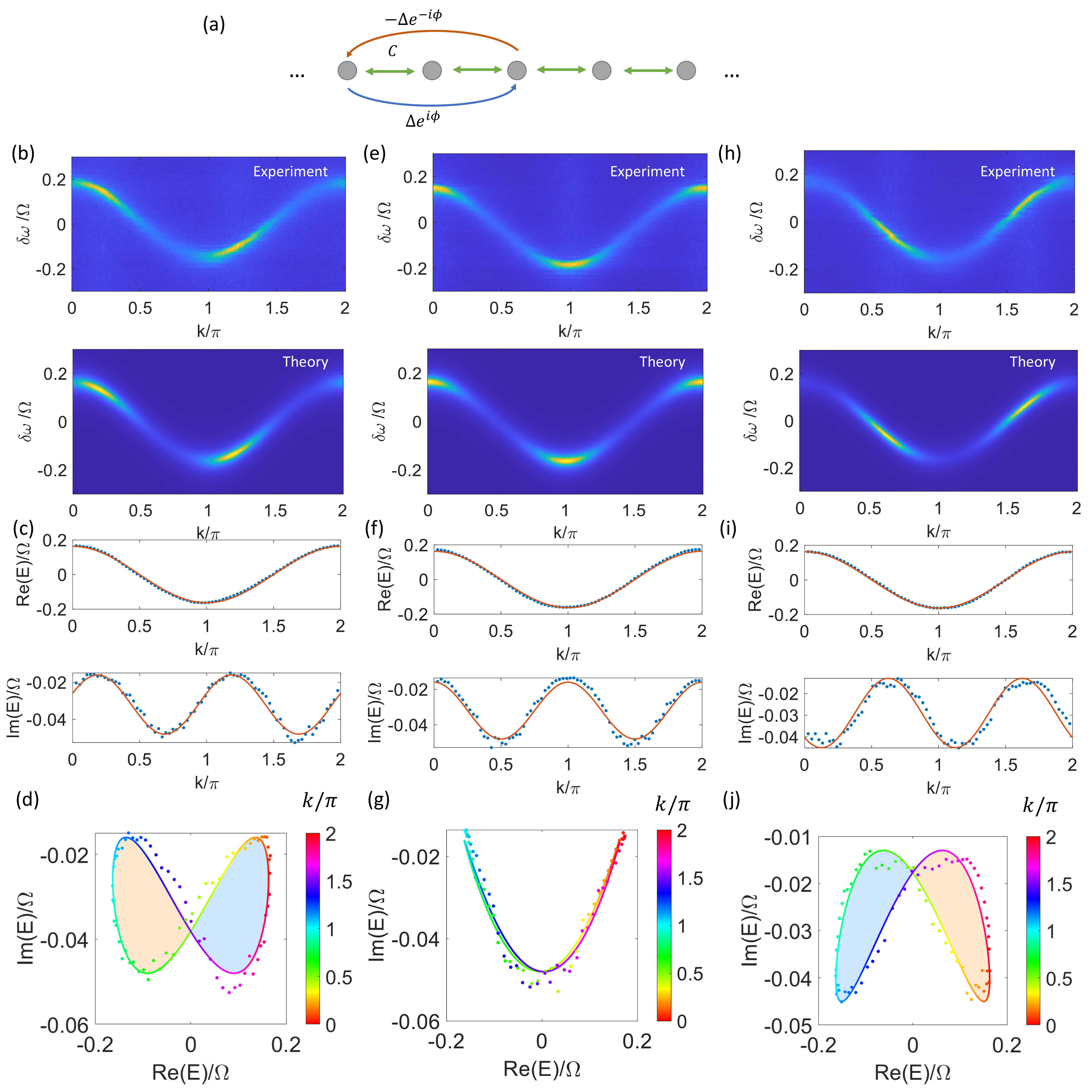}
    \caption{Experimental examples for a lattice with 1st-order Hermitian coupling and 2nd-order skew-Hermitian coupling. (a) Lattice sketch. (b--d) $\phi=\pi/8$. (e--g) $\phi=\pi/2$. (h--j) $\phi=-3\pi/4$. }
    \label{fig:ratio_1_2}
\end{figure}

\begin{figure}[t]
    \centering
    \includegraphics[width=0.9\columnwidth]{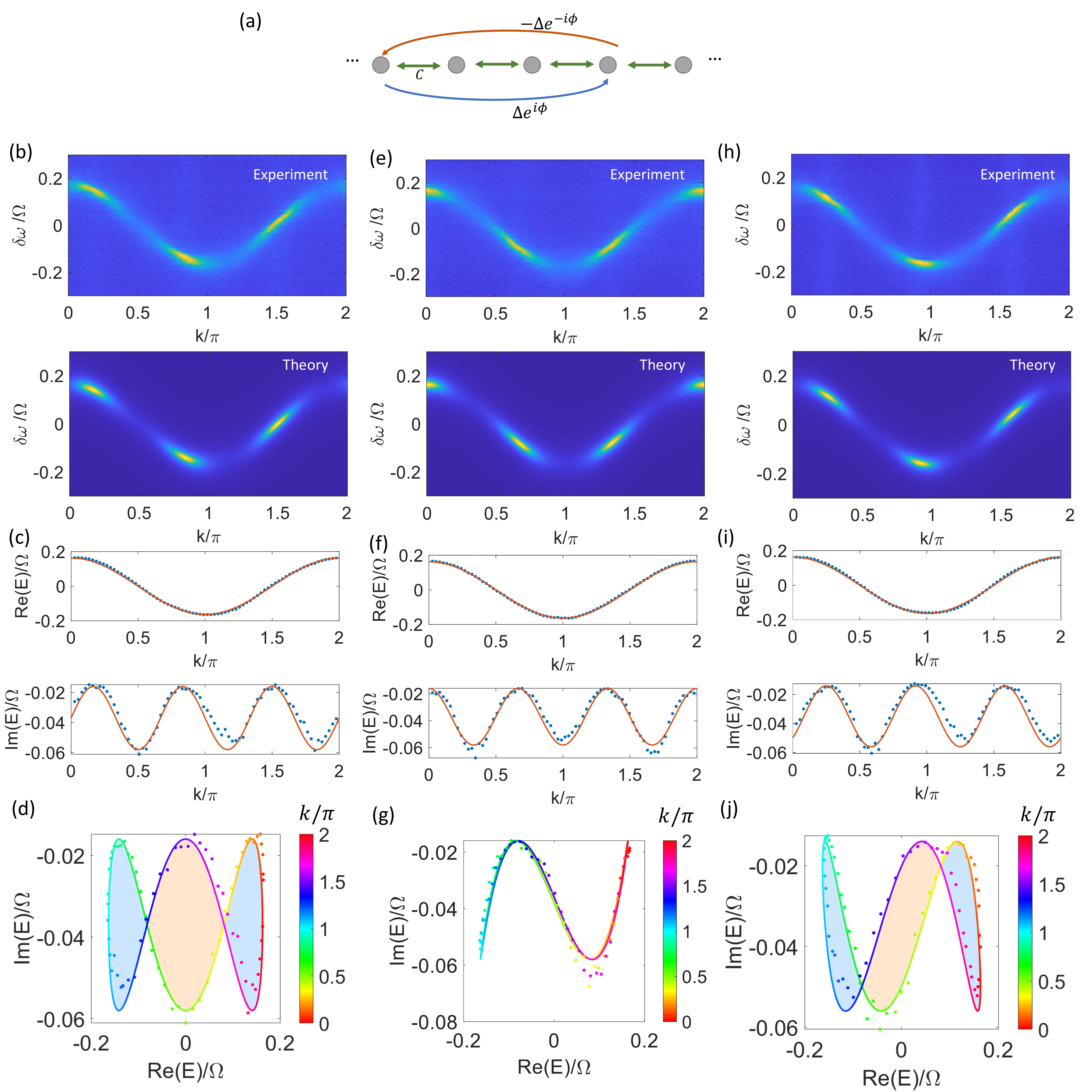}
    \caption{Experimental examples for a lattice with 1st-order Hermitian coupling and 3rd-order skew-Hermitian coupling. (a) Lattice sketch. (b--d) $\phi=0$. (e--g) $\phi=\pi/2$. (h--j) $\phi=-\pi/4$. }
    \label{fig:ratio_1_3}
\end{figure}

\begin{figure}[t]
    \centering
    \includegraphics[width=0.9\columnwidth]{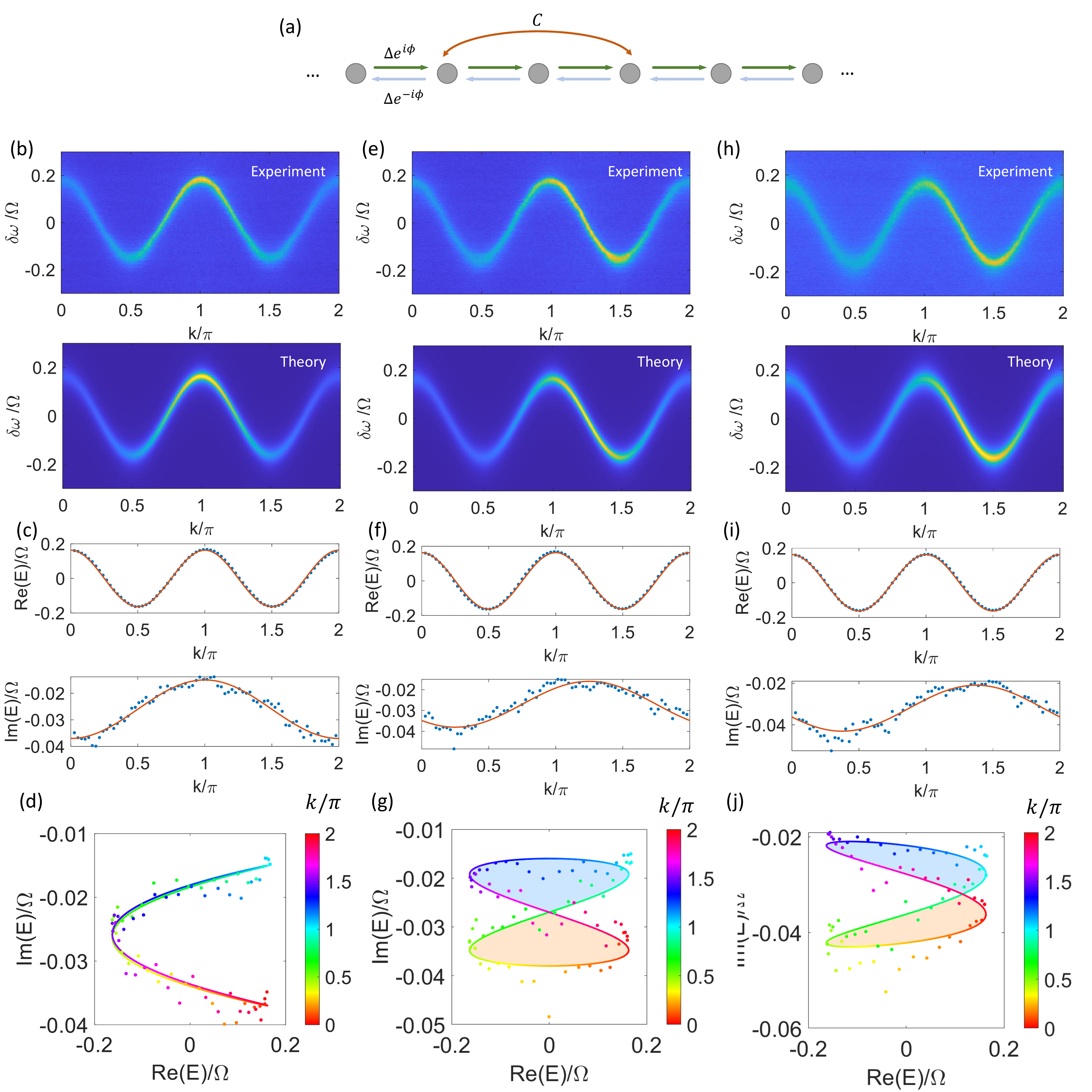}
    \caption{Experimental examples for a lattice with 2nd-order Hermitian coupling and 1st-order skew-Hermitian coupling. (a) Lattice sketch. (b--d) $\phi=-\pi/2$. (e--g) $\phi=-3\pi/4$. (h--j) $\phi=-7\pi/8$.}
    \label{fig:ratio_2_1}
\end{figure}

\begin{figure}[t]
    \centering
    \includegraphics[width=0.9\columnwidth]{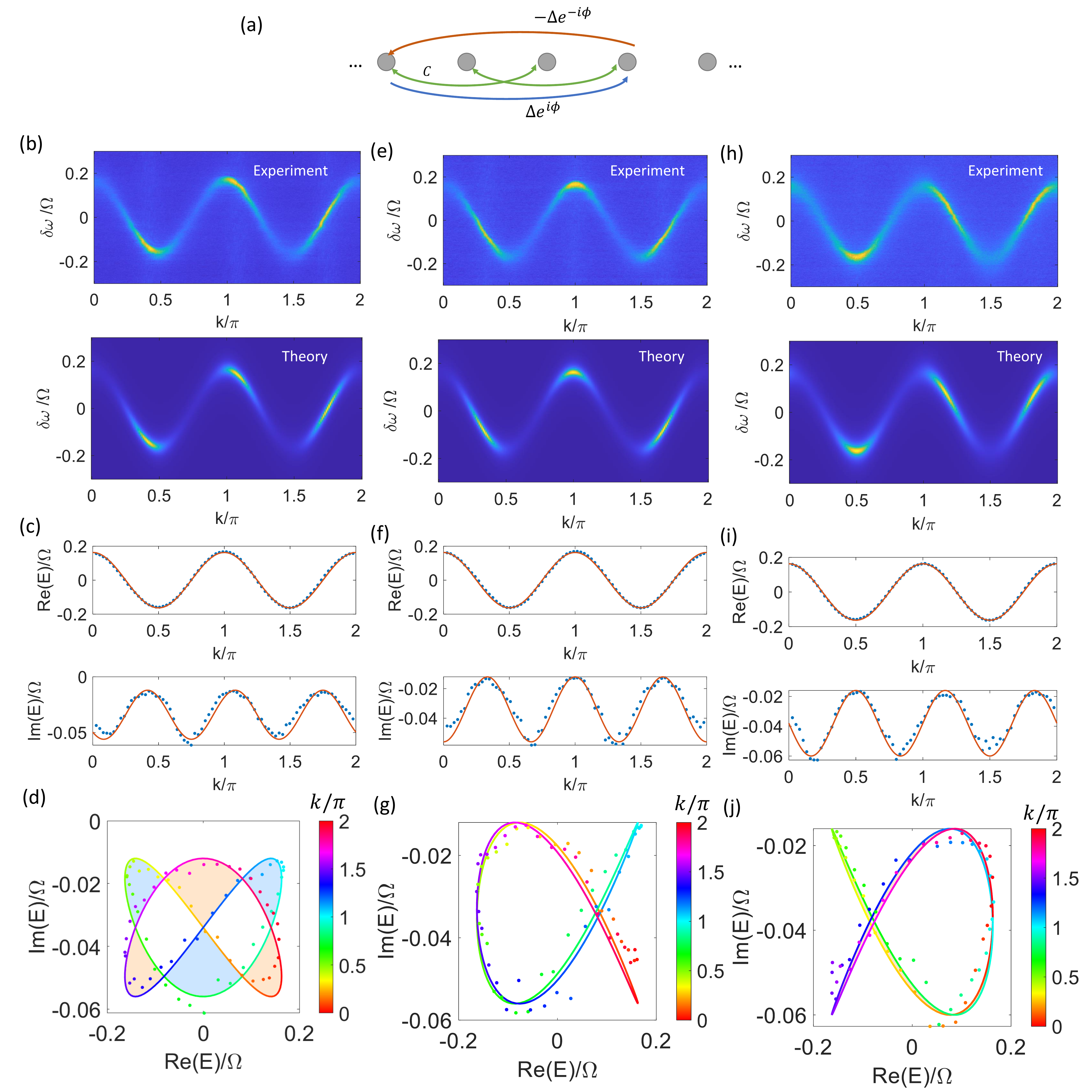}
    \caption{Experimental examples for a lattice with 2nd-order Hermitian coupling and 3rd-order skew-Hermitian coupling. (a) Lattice sketch. (b--d) $\phi=-3\pi/4$. (e--g) $\phi=-\pi/2$. (h--j) $\phi=\pi$.}
    \label{fig:ratio_2_3}
\end{figure}

\begin{figure}[t]
    \centering
    \includegraphics[width=0.75\columnwidth]{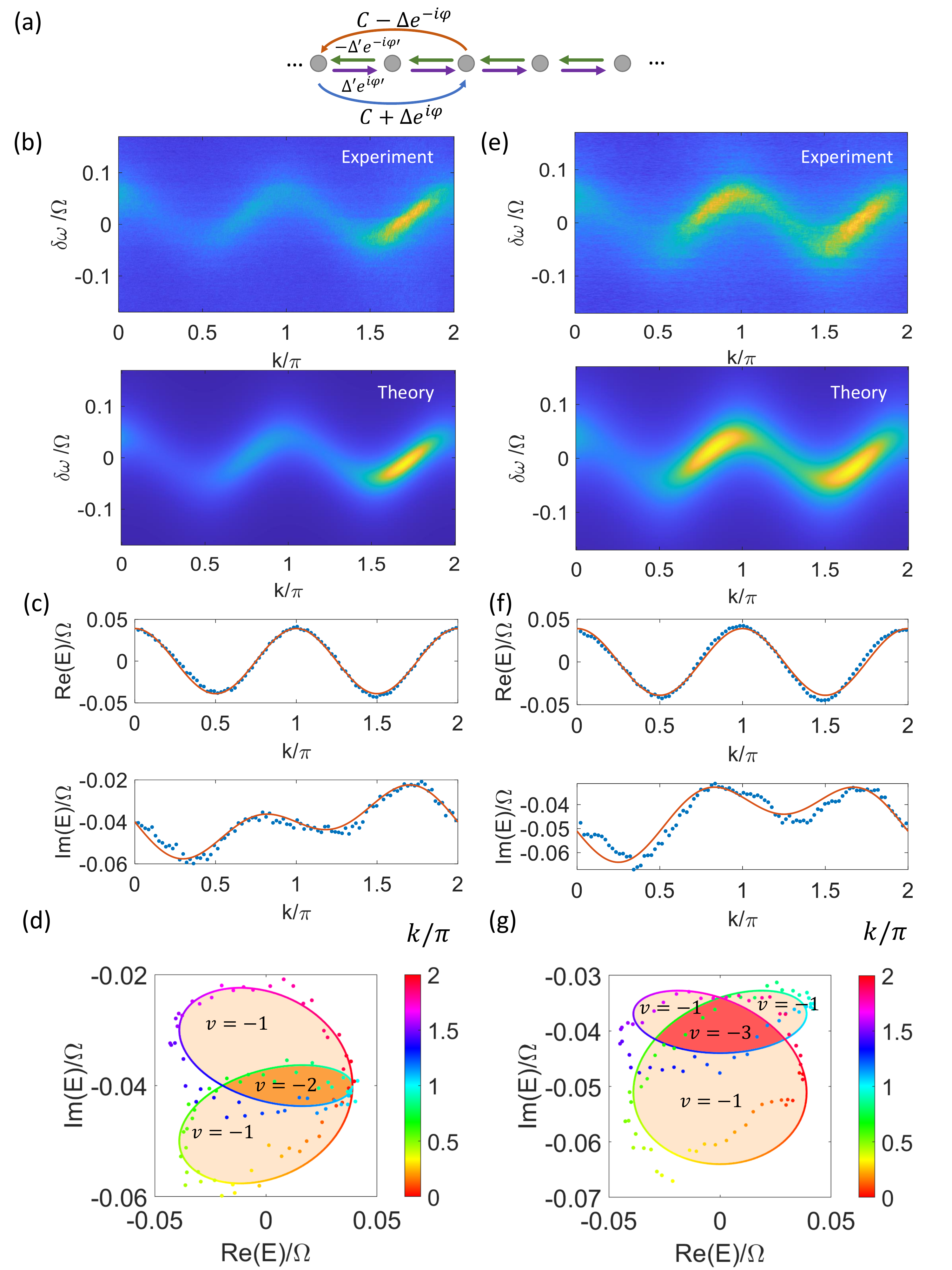}
    \caption{Experimental examples for a lattice with 1st-order skew-Hermitian coupling and 2nd-order non-Hermitian coupling. (a) Lattice sketch. (b--d) $\phi=\phi'=\pi$. (e--g) $\phi=\pi$, $\phi'=-3\pi/4$.}
    \label{fig:complex}
\end{figure}

In Figs.~2(d,g,h) of the main manuscript, we have shown an example of a lattice with 1st-order Hermitian coupling and 2nd-order skew-Hermitian coupling. In this case, we have a dispersion 
\begin{equation}
E(k)=2C \cos (k)+2i\Delta \sin (2k+\phi)-i\gamma,    
\end{equation}
with $2C= 0.163\ \Omega$ and $2\Delta = 0.016\ \Omega$ and $\phi\approx 0$. Here in Fig.~\ref{fig:ratio_1_2} we show another three instances for the same values of $C$ and $\Delta$ yet different values of $\phi$. The results in Figs.~\ref{fig:ratio_1_2}(b--d) correspond to the case $\phi=\pi/8$, which gives a butterfly-like winding pattern with winding number $v=-1$ (light orange shaded) and $v=1$ (light blue shaded). By contrast, the results in Figs.~\ref{fig:ratio_1_2}(e--g) have $\phi=\pi/2$. In this case, the topology is trivial and the energy band instead forms a curve. In order to see the whole variations of $k$ along the curve in Figs.~\ref{fig:ratio_1_2} (g), we deliberately shift the theoretical curve to $\phi=\pi/2-0.05$. In Figs.~\ref{fig:ratio_1_2}(h--j), we show the case with $\phi = -3\pi/4$. The winding pattern in Figs.~\ref{fig:ratio_1_2}(j) is approximately an up-side-down version of the pattern in Figs.~\ref{fig:ratio_1_2}(d).

Then, we show some further examples for the lattice in Figs.~4(a--c) of the main manuscript with 1st-order Hermitian coupling and 3rd-order skew-Hermitian coupling. For convenience, we replicate this instance in Figs.~\ref{fig:ratio_1_3}(b,d), which additionally includes the $\mathrm{Re}(E),\mathrm{Im}(E)-k$ plots as shown in Fig.~\ref{fig:ratio_1_3}(c). Such a lattice has a dispersion 
\begin{equation}
E(k)=2C \cos (k)+2i\Delta \sin (3k+\phi)-i\gamma,
\end{equation}
with $2C= 0.163\ \Omega$ and $2\Delta = 0.021\ \Omega$ and $\phi= 0$. Then, we keep the values of $C$ and $\Delta$ and vary $\phi$. A $\phi=\pi/2$ example is shown in Figs.~\ref{fig:ratio_1_3}(e--g). This is a topologically trivial case since the energy band forms a curve instead of a loop, see Fig.~\ref{fig:ratio_1_3}(g). The theoretical prediction (curve) is also deliberately shifted to $\phi=\pi/2-0.05$. Another non-trivial case is shown in Figs.~\ref{fig:ratio_1_3}(h--j) with $\phi=-\pi/4$, where there are three enclosed areas with nonzero winding numbers, from left to right, $v=1,-1,1$, respectively.

Next, we show an example with 2nd-order Hermitian coupling and 1st-order skew-Hermitian coupling, with the lattice sketched in Fig.~\ref{fig:ratio_2_1}(a). Such a lattice has a dispersion 
\begin{equation}
E(k)=2C \cos (2k)+2i\Delta \sin (k+\phi)-i\gamma,
\end{equation}
where we use $2C= 0.163\ \Omega$ and $2\Delta = 0.011\ \Omega$. Here we show two example cases with different values of $\phi$. Figs.~\ref{fig:ratio_2_1}(b--d) correspond to the case with $\phi=-\pi/2$, where the energy band becomes topologically trivial forming a curve. To clearly show the $k$-dependent nature, in Fig.~\ref{fig:ratio_2_1}(d) we deliberately shift the theoretical curve to $\phi=-\pi/2-0.02$. The results shown in Figs.~\ref{fig:ratio_2_1}(e--g) have a phase $\phi=-3\pi/4$, where the winding pattern looks like a bow-tie rotated by 90 degree, with the winding number $v=1,-1$ from top to bottom, respectively. A slightly different case can be seen when we further change the phase to $\phi=-7\pi/8$, as shown in Figs.~\ref{fig:ratio_2_1}(h--j).

In Fig.~4(d) of the main manuscript, we showed a lattice with 2nd-order Hermitian coupling and 3rd-order skew-Hermitian coupling. The example shown there (Figs.~4(e,f) in the main text) has the dispersion
\begin{equation}
E(k)=2C \cos (2k)+2i\Delta \sin (3k+\phi)-i\gamma,
\end{equation}
where $2C= 0.163\ \Omega$, $2\Delta = 0.022\ \Omega$ and $\phi=-3\pi/4$. For convenience we replicate the lattice sketch in Fig.~\ref{fig:ratio_2_3}(a) and the same experimental plots in Figs.~\ref{fig:ratio_2_3}(b--d) while adding the $k$-dependent band energy plots in Fig.~\ref{fig:ratio_2_3}(c). Now we show two more instances with other values of $\phi$. One case is shown in Figs.~\ref{fig:ratio_2_3}(e--g), where we have $\phi=-\pi/2$ and the band winds into a $\alpha$-like shaped curve. Here in the plot of Fig.~\ref{fig:ratio_2_3}(g) we use a slightly deviated theory curve with $\phi=-\pi/2-0.05$ to better show the orientation of the winding. In Figs.~\ref{fig:ratio_2_3}(h--j) we show the case with $\phi=\pi$, where the band winds into a horizontally flipped $\alpha$-shape. Similarly, we deliberate shift the theory curve with a $\phi=\pi-0.05$ to show the complete $k$-dependency.

Finally, we show a more complicated lattice model in Fig.~\ref{fig:complex}(a). This model has been shown in Fig.~4(g) of the main manuscript, where there are 1st-order skew-Hermitian coupling and a 2nd-order non-Hermitian coupling with both Hermitian and skew-Hermitian parts. This lattice has the band energy
\begin{equation}
E(k)=2C \cos (2k)+2i\Delta' \sin (k+\phi')+2i\Delta \sin (2k+\phi)-i\gamma,
\end{equation}
where in the example shown in Figs.~4(h,i) of the main manuscript the parameters are $2C=0.039\ \Omega$, $2\Delta=2\Delta'=0.010\ \Omega$, and $\phi=\phi'=\pi$. We have replicated the plots of this example in Figs.~\ref{fig:complex}(b--d) while adding the $k$-resolved band energy plots in Fig.~\ref{fig:complex}(c). Additionally, here we show another example by changing the phases to $\phi=\pi$ and $\phi'=-3\pi/4$ in Figs.~\ref{fig:complex}(e--g). The deviation between the experimentally obtained $\mathrm{Im}(E)$ and the theoretical predictions in Figs.~\ref{fig:complex}(f,g) can be explained by a small ($\sim -0.15\pi$) possible phase drift in $\phi$, which can be relieved through more careful calibrations of the RF amplifiers. In this example, the band winds into a more complicated loop dividing into four regions with $v=-3$ in the red-shaded region and $v=-1$ in the light-orange-shaded regions of Fig.~\ref{fig:complex}(g).

\newpage
\pagebreak

\end{document}